\pdfoutput=1

\documentclass[12pt,a4paper]{article}

\usepackage{ifthen} 
\newboolean{pdflatex}
\setboolean{pdflatex}{true} 

\newboolean{articletitles}
\setboolean{articletitles}{true} 

\newboolean{uprightparticles}
\setboolean{uprightparticles}{true} 

\def\paperauthors{O. Lupton, M. Vesterinen} 
\def\paperasciititle{Simultaneously determining the W boson mass and parton shower model parameters}
\def\papertitle{Simultaneously determining the \\ \Wpm boson mass and parton \\ shower model parameters}
\def\paperkeywords{{High Energy Physics}, {LHCb}, {W boson}} 
\def\papercopyright{\paperauthors}
\def\paperlicenceurl{https://creativecommons.org/licenses/by/4.0/}

\usepackage[top=1in, bottom=1.25in, left=1in, right=1in]{geometry}

\usepackage{microtype}
\usepackage{lineno}  
\usepackage{xspace} 
\usepackage{caption} 

\usepackage{graphicx}  
\usepackage{color}
\usepackage{colortbl}
\graphicspath{{./figs/}} 
\DeclareGraphicsExtensions{.pdf,.PDF,png,.PNG}

\usepackage{amsmath} 
\usepackage{amssymb}
\usepackage{amsfonts}
\usepackage{upgreek} 

\newcommand*\patchAmsMathEnvironmentForLineno[1]{%
\expandafter\let\csname old#1\expandafter\endcsname\csname #1\endcsname
\expandafter\let\csname oldend#1\expandafter\endcsname\csname
end#1\endcsname
 \renewenvironment{#1}%
   {\linenomath\csname old#1\endcsname}%
   {\csname oldend#1\endcsname\endlinenomath}%
}
\newcommand*\patchBothAmsMathEnvironmentsForLineno[1]{%
  \patchAmsMathEnvironmentForLineno{#1}%
  \patchAmsMathEnvironmentForLineno{#1*}%
}
\AtBeginDocument{%
\patchBothAmsMathEnvironmentsForLineno{equation}%
\patchBothAmsMathEnvironmentsForLineno{align}%
\patchBothAmsMathEnvironmentsForLineno{flalign}%
\patchBothAmsMathEnvironmentsForLineno{alignat}%
\patchBothAmsMathEnvironmentsForLineno{gather}%
\patchBothAmsMathEnvironmentsForLineno{multline}%
\patchBothAmsMathEnvironmentsForLineno{eqnarray}%
}

\usepackage{hyperxmp}

\usepackage[pdftex,
            pdfauthor={\paperauthors},
            pdftitle={\paperasciititle},
            pdfkeywords={\paperkeywords},
            pdfcopyright={Copyright (C) \papercopyright},
            pdflicenseurl={\paperlicenceurl}]{hyperref}

\usepackage[colorinlistoftodos,textsize=scriptsize]{todonotes}

\usepackage[all]{hypcap} 

\usepackage{xspace} 
\usepackage{upgreek}

\def\lhcb   {\mbox{LHCb}\xspace}
\def\atlas  {\mbox{ATLAS}\xspace}

\def\cdf    {\mbox{CDF}\xspace}
\def\dzero  {\mbox{D0}\xspace}

\def\lhc    {\mbox{LHC}\xspace}

\def\MagUp {\mbox{\em Mag\kern -0.05em Up}\xspace}

\ifthenelse{\boolean{uprightparticles}}%
{

 \def\Pmu         {\ensuremath{\upmu}\xspace}                 
 \def\Pnu         {\ensuremath{\upnu}\xspace}

 \def\PDelta      {\ensuremath{\Delta}\xspace}                 
 \def\PXi         {\ensuremath{\Xi}\xspace}                 
 \def\PLambda     {\ensuremath{\Lambda}\xspace}                 
 \def\PSigma      {\ensuremath{\Sigma}\xspace}                 
 \def\POmega      {\ensuremath{\Omega}\xspace}                 
 \def\PUpsilon    {\ensuremath{\Upsilon}\xspace}

 \def\PB      {\ensuremath{\mathrm{B}}\xspace}                 
                  
 \def\PD      {\ensuremath{\mathrm{D}}\xspace}

 \def\PK      {\ensuremath{\mathrm{K}}\xspace}

 \def\PV      {\ensuremath{\mathrm{V}}\xspace}                 
 \def\PW      {\ensuremath{\mathrm{W}}\xspace}

 \def\PZ      {\ensuremath{\mathrm{Z}}\xspace}

 \def\Pi      {\ensuremath{\mathrm{i}}\xspace}

 \def\Pp      {\ensuremath{\mathrm{p}}\xspace}

 \def\Ps      {\ensuremath{\mathrm{s}}\xspace}

 \def\thebaroffset{0.0em}
}
{

 \def\Pmu         {\ensuremath{\mu}\xspace}                 
 \def\Pnu         {\ensuremath{\nu}\xspace}

 \mathchardef\PDelta="7101
 \mathchardef\PXi="7104
 \mathchardef\PLambda="7103
 \mathchardef\PSigma="7106
 \mathchardef\POmega="710A
 \mathchardef\PUpsilon="7107
                  
 \def\PB      {\ensuremath{B}\xspace}                 
                  
 \def\PD      {\ensuremath{D}\xspace}

 \def\PK      {\ensuremath{K}\xspace}

 \def\PV      {\ensuremath{V}\xspace}                 
 \def\PW      {\ensuremath{W}\xspace}

 \def\PZ      {\ensuremath{Z}\xspace}

 \def\Pi      {\ensuremath{i}\xspace}

 \def\Pp      {\ensuremath{p}\xspace}

 \def\Ps      {\ensuremath{s}\xspace}

 \def\thebaroffset{0.18em}
}
\newcommand{\offsetoverline}[2][\thebaroffset]{\kern #1\overline{\kern -#1 #2}}%

\makeatletter
\ifcase \@ptsize \relax
  \newcommand{\miniscule}{\@setfontsize\miniscule{4}{5}}
\or
  \newcommand{\miniscule}{\@setfontsize\miniscule{5}{6}}
\or
  \newcommand{\miniscule}{\@setfontsize\miniscule{5}{6}}
\fi
\makeatother

\DeclareRobustCommand{\optbar}[1]{\shortstack{{\miniscule (\rule[.5ex]{1.25em}{.18mm})}
  \\ [-.7ex] $#1$}}

\def\muon       {{\ensuremath{\Pmu}}\xspace}

\def\Wp     {{\ensuremath{\PW^+}}\xspace}
\def\Wm     {{\ensuremath{\PW^-}}\xspace}
\def\Wpm    {{\ensuremath{\PW^\pm}}\xspace}

\def\Zz     {{\ensuremath{\PZ^0}}\xspace}

\def\squark    {{\ensuremath{\Ps}}\xspace}

\def\KorKbar {\kern \thebaroffset\optbar{\kern -\thebaroffset \PK}{}\xspace}

\def\DorDbar {\kern \thebaroffset\optbar{\kern -\thebaroffset \PD}\xspace}

\def\B       {{\ensuremath{\PB}}\xspace}

\def\BorBbar {\kern \thebaroffset\optbar{\kern -\thebaroffset \PB}\xspace}

\def\Bd      {{\ensuremath{\B^0}}\xspace}

\def\BdorBdbar {\kern \thebaroffset\optbar{\kern -\thebaroffset \Bd}\xspace}

\def\Bs      {{\ensuremath{\B^0_\squark}}\xspace}

\def\BsorBsbar {\kern \thebaroffset\optbar{\kern -\thebaroffset \Bs}\xspace}

\def\Y#1S{\ensuremath{\PUpsilon{(#1S)}}\xspace}

\def\proton      {{\ensuremath{\Pp}}\xspace}

\def\LorLbar     {\kern \thebaroffset\optbar{\kern -\thebaroffset \PLambda}\xspace}

\newcommand{\decay}[2]{\ensuremath{#1\!\to #2}\xspace}         

\def\to                 {\ensuremath{\rightarrow}\xspace}

\newcommand{\mW}{{\ensuremath{m_{\PW}}}\xspace}

\newcommand{\as}{{\ensuremath{\alpha_s}}\xspace}

\def\AT#1     {\ensuremath{A_{\mathrm{T}}^{#1}}\xspace}           

\def\C#1      {\ensuremath{\mathcal{C}_{#1}}\xspace}                       
\def\Cp#1     {\ensuremath{\mathcal{C}_{#1}^{'}}\xspace}                    
\def\Ceff#1   {\ensuremath{\mathcal{C}_{#1}^{\mathrm{(eff)}}}\xspace}        
\def\Cpeff#1  {\ensuremath{\mathcal{C}_{#1}^{'\mathrm{(eff)}}}\xspace}       
\def\Ope#1    {\ensuremath{\mathcal{O}_{#1}}\xspace}                       
\def\Opep#1   {\ensuremath{\mathcal{O}_{#1}^{'}}\xspace}                    

\newcommand{\aunit}[1]{\ensuremath{\text{\,#1}}}       

\newcommand{\tev}{\aunit{Te\kern -0.1em V}\xspace}
\newcommand{\gev}{\aunit{Ge\kern -0.1em V}\xspace}
\newcommand{\mev}{\aunit{Me\kern -0.1em V}\xspace}
\newcommand{\kev}{\aunit{ke\kern -0.1em V}\xspace}
\newcommand{\ev}{\aunit{e\kern -0.1em V}\xspace}
\newcommand{\mevc}{\ensuremath{\aunit{Me\kern -0.1em V\!/}c}\xspace}
\newcommand{\gevc}{\ensuremath{\aunit{Ge\kern -0.1em V\!/}c}\xspace}
\newcommand{\mevcc}{\ensuremath{\aunit{Me\kern -0.1em V\!/}c^2}\xspace}
\newcommand{\gevcc}{\ensuremath{\aunit{Ge\kern -0.1em V\!/}c^2}\xspace}

\def\fb   {\ensuremath{\aunit{fb}}\xspace}
\def\invfb   {\ensuremath{\fb^{-1}}\xspace}

\def\gsim{{~\raise.15em\hbox{$>$}\kern-.85em
          \lower.35em\hbox{$\sim$}~}\xspace}
\def\lsim{{~\raise.15em\hbox{$<$}\kern-.85em
          \lower.35em\hbox{$\sim$}~}\xspace}

\def\sqs   {\ensuremath{\protect\sqrt{s}}\xspace}

\def\pt         {\ensuremath{p_{\mathrm{T}}}\xspace}
\def\ptV        {\ensuremath{p_{\mathrm{T}}^{V}}\xspace}
\def\ptW        {\ensuremath{p_{\mathrm{T}}^{\PW}}\xspace}
\def\ptZ        {\ensuremath{p_{\mathrm{T}}^{\PZ}}\xspace}
\def\ptl        {\ensuremath{p_{\mathrm{T}}^{\ell}}\xspace}
\def\ptmu       {\ensuremath{p_{\mathrm{T}}^{\Pmu}}\xspace}
\def\ptnu       {\ensuremath{p_{\mathrm{T}}^{\Pnu}}\xspace}

\def\kt         {\ensuremath{k_{\mathrm{T}}}\xspace}

\def\pythia     {\mbox{\textsc{Pythia}}\xspace}

\xspace

\def\tell1  {TELL1\xspace}
\def\ukl1   {UKL1\xspace}

\newcommand{\ie}{\mbox{\itshape i.e.}\xspace}

\newcommand{\IKT}{{\ensuremath{k_{\mathrm{T}}^{\text{intr.}}}}\xspace}

\newcommand{\IKTm}{{\ensuremath{k_{\mathrm{T}, -}^{\text{intr.}}}}\xspace}

\newcommand{\asp}{{\ensuremath{\alpha_{s}^{+}}}\xspace}
\newcommand{\asm}{{\ensuremath{\alpha_{s}^{-}}}\xspace}
\newcommand{\aspm}{{\ensuremath{\alpha_{s}^{\pm}}}\xspace}
\newcommand{\ptmin}{{\ensuremath{p_{\mathrm{T}}^{\text{min.}}}}\xspace}
\newcommand{\ptmax}{{\ensuremath{p_{\mathrm{T}}^{\text{max.}}}}\xspace} 

\usepackage{cite} 
\usepackage{mciteplus}
\begin{document}
\renewcommand{\thefootnote}{\fnsymbol{footnote}}
\setcounter{footnote}{1}


\begin{titlepage}
\pagenumbering{roman}

\noindent
\vspace*{5.0cm}

{\normalfont\bfseries\boldmath\huge
\begin{center}
  \papertitle 
\end{center}
}

\vspace*{2.0cm}

\begin{center}
  \paperauthors\\[12pt]
  \textit{University of Warwick, Coventry, United Kingdom}
\end{center}

\vspace*{2cm}

\begin{abstract}
  \noindent
  We explore the possibility of simultaneously determining the \PW boson mass, \mW, and QCD-related
  nuisance parameters that affect the \PW boson \pt spectrum from a fit to the \pt spectrum of the
  muon in the leptonic decay \decay{\PW}{\muon\Pnu}.
  The study is performed using pseudodata generated using the parton shower event generator \pythia and the muon is required to fall
  in a kinematic region corresponding to the approximate acceptance of the \lhcb detector.
  We find that the proposed method performs well and has little trouble disentangling variations in
  the muon \pt spectrum due to \mW  from those due to the \PW boson \pt model.
\end{abstract}
\end{titlepage}

\renewcommand{\thefootnote}{\arabic{footnote}}
\setcounter{footnote}{0}



\pagestyle{plain} 
\setcounter{page}{1}
\pagenumbering{arabic}

\section{Introduction}
\label{sec:Introduction}
Global fits to precision electroweak observables are a powerful probe of physics
beyond the Standard Model (SM).
One input to these fits, the \PW boson mass, \mW, is of particular interest
because it is determined indirectly by the electroweak fits more precisely than
it has been measured directly.
The recent Gfitter electroweak fit update~\cite{Haller:2018nnx} indirectly
determines $\mW = 80.354\pm0.007\gevcc$, while the latest average of direct 
measurements, which is dominated by inputs from the  
\cdf~\cite{Aaltonen:2012bp}, \dzero~\cite{Abazov:2012bv} and 
\atlas~\cite{Aaboud:2017svj} collaborations, is
$\mW = 80.379\pm0.012\gevcc$~\cite{PDG2018}.
Improving the precision of the direct measurement is therefore well motivated.

Measurements of \mW at hadron colliders have to date been based on three  
different observables in \decay{\PW}{\ell\nu} decays, where $\ell$ represents
an electron or muon. These are: the transverse momentum of the charged lepton,
\ptl, the missing transverse momentum, \ptnu, and the transverse mass
$m_{T} = \sqrt{2\ptl\ptnu\left(1 - \cos\Delta\phi\right)}$, where $\Delta\phi$ is
the opening angle between the charged and neutral lepton momenta in the plane
transverse to the beams.
There are two, closely related, sources of systematic uncertainty that potentially
limit the precision with which \mW can be measured at the \lhc.
The first is the parton distribution functions (PDFs) that primarily determine the
rapidity, $y$, distribution of the \PW bosons.
The second is the transverse momentum distribution of the \PW bosons, \ptW.
The \ptl distribution is particularly sensitive to the latter.

It has previously been suggested that a measurement of \mW in the forward 
kinematic region covered by the \lhcb experiment would be of particular interest
due to the predicted anticorrelation of PDF uncertainties between  measurements in
the central and forward rapidity regions~\cite{Bozzi:2015zja}.
Further studies of the PDF uncertainties affecting an \lhcb measurement of \mW
have been developed in Ref.~\cite{Farry:2019rfg}, including suggestions of how
these can be reduced by using in-situ constraints.
Since the proposed \lhcb measurement of \mW is based on the \ptl spectrum, it is
particularly susceptible to uncertainties in the \ptW spectrum.
Our attention is therefore drawn to mitigating strategies for that source of uncertainty in the 
context of an \lhcb measurement.

Fixed order QCD corrections to the \Wpm and \Zz cross sections are known
fully  differentially up to $\mathcal{O}\left(\alpha_{s}^2\right)$~\cite{%
Hamberg:1990np,vanNeerven:1991gh,%
Catani:2009sm,Gavin:2010az,Anastasiou:2003ds}, and
calculations differential in the gauge boson transverse momentum, \ptV, have 
recently been made up to 
$\mathcal{O}\left(\alpha_{s}^{3}\right)$~\cite{%
Boughezal:2015dva,Gehrmann-DeRidder:2017mvr}.
Electroweak corrections are known up to next-to-leading order~\cite{Dittmaier:2001ay,Arbuzov:2005dd,CarloniCalame:2006zq,Baur:2004ig,Barze:2012tt}.
These state-of-the-art fixed order calculations are crucial, and the higher order
corrections are important at larger \ptV values.
The bulk of the \ptV distribution is, however, situated in the $\ptV\lesssim 
m_{\PV}$ region, where large logarithmic terms must be resummed to achieve an 
accurate prediction.
This can be approached in two ways.
The first is to use analytic resummation techniques, where next-to-next-to-leading
logarithmic accuracy (NNLL) is well 
known~\cite{Becher:2011xn,Bozzi:2010xn,Banfi:2012du,Alioli:2015toa,Coradeschi:2017zzw,Camarda:2019zyx} and
N$^3$LL~\cite{Bizon:2018foh} has recently been achieved.
The second approach is to use parton-shower algorithms, such as Herwig++~\cite{Bellm:2015jjp},
Pythia~\cite{Sjostrand:2006za,*Sjostrand:2007gs} and Sherpa~\cite{Gleisberg:2008ta}.

While the improvements in calculations of the \ptW spectrum in recent years are
impressive, the precision of the state-of-the-art calculations is yet to reach the
$\mathcal{O}\left(1\%\right)$ level required for a $\mathcal{O}(10\mevcc)$ 
measurement of \mW at the \lhc.
One approach to determining the \ptW spectrum with this precision is to study the
\ptZ spectrum, which can be measured extremely precisely in the regions of phase
space that are likely to produce two final-state leptons in the relevant detector
acceptance, and use these measurements to infer the \ptW spectrum with, ideally,
reduced uncertainties with respect to the direct calculation of \ptW.
How best to evaluate robust theoretical uncertainties in this approach is an
open topic.
Independently of whether explicit constraints from \Zz data are included in 
experimental fits of \mW, it is important to define well-motivated nuisance
parameters that can be varied during the experimental analyses.

This paper explores the possibility of simultaneously determining \mW and
nuisance parameters relating to \ptW in the context of the proposed measurement of
\mW at \lhcb using the muon transverse momentum spectrum, \ptmu.
This study identifies two parameters of the 
\pythia~\cite{Sjostrand:2006za,*Sjostrand:2007gs} Monte Carlo generator, which 
strongly affect the \ptW distribution, as examples of nuisance parameters that could
be varied in an \mW measurement~\cite{Skands:2010ak,Skands:2014pea}.
One of these is related to the intrinsic parton \kt, and the second is related to
the strong coupling constant.
The ATLAS collaboration also varied the intrinsic $k_T$ cut off parameter in the \texttt{AZ} tune of Pythia~\cite{Aad:2014xaa},
but this parameter is found to be far less influential than the two parameters that we consider.
It is, of course, unlikely that \pythia with only these two nuisance parameters
would have sufficient freedom to describe \ptW sufficiently accurately in a real 
measurement of \mW, and they are unlikely to accurately reflect the residual
perturbative uncertainties in state-of-the-art calculations of \ptW.
It is nonetheless interesting to consider the \ptmu fit performance with this
simplified setup, with the expectation that in a real measurement of \mW a tool 
with higher formal accuracy would be used in place of \pythia, and the nuisance 
parameters used in the \ptmu fit would be chosen to -- as far as possible -- 
reflect the residual uncertainty on \ptW after, for example, tuning using \ptZ
and other data.

The possibility of determining these parameters directly from the \PW boson data
is an attractive one, as it could allow a measurement of \mW to reduce its
sensitivity to imperfectly modelled differences between \Wpm and \Zz production,
such as heavy quark effects~\cite{Pietrulewicz:2017gxc,Bagnaschi:2018dnh} and
flavour-dependent parton transverse momenta~\cite{Bacchetta:2018lna}, and avoid
constraining nuisance parameters to values determined using measurements of \Zz production.
The differences between \Wpm and \Zz production, and the associated uncertainties in a measurement of \mW,
were extensively studied by the ATLAS collaboration~\cite{ATL-PHYS-PUB-2014-015}.

\section{\boldmath Simulation of \PW production and reweighting}
\begin{figure}
    \centering
    \includegraphics{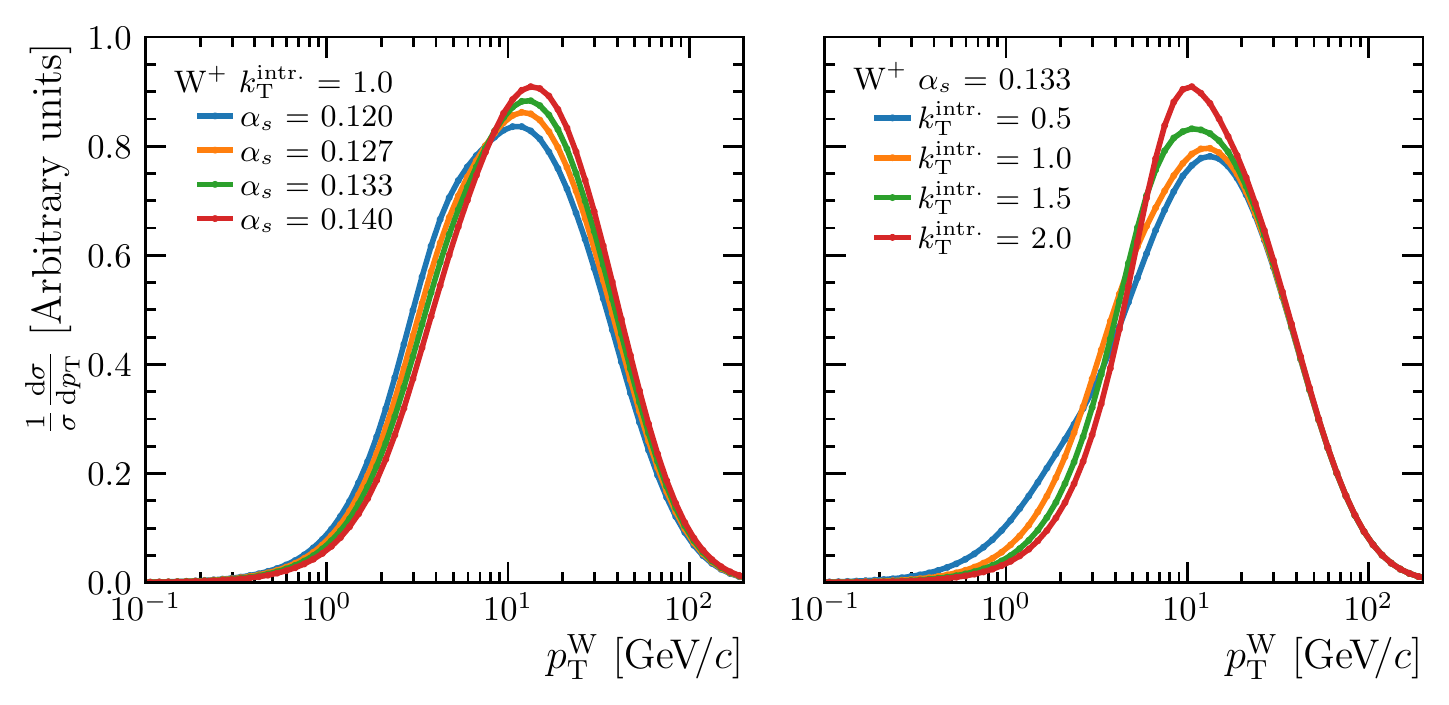}
    \caption{Illustration of variations in the \Wp boson \pt spectrum corresponding
             to variations in the \as (left) and \IKT (right) nuisance parameters.
             No kinematic requirements have been placed on the \Wp decay products.
             The equivalent distributions for the \Wm, which qualitatively look very
             similar, are shown in Appendix~\ref{app:extra_massfit_plots}.}
    \label{fig:Wpt_spectrum_Wp}
\end{figure}
Monte Carlo events of the inclusive process $\proton\proton\to\PW\to\muon\Pnu$, at a
centre-of-mass energy $\sqs = 13\tev$, are generated using 
\pythia~\cite{Sjostrand:2006za,*Sjostrand:2007gs} version 8.235 and the 
\texttt{NNPDF23\_lo\_as\_0130\_qed} PDF set~\cite{Ball:2013hta}.
Samples are generated for a $4 \times 4$ grid of different parton-shower \as
and \IKT parameters.
These are the two parameters, in \pythia, that most strongly affect the \ptW distribution.
Their precise definitions, and the ranges over which they are varied, are
detailed in Appendix~\ref{app:pythia_params}.
For this study around $1.7\times10^{8}$ events are produced at each of the 16 grid points,
corresponding to around three times the expected yields given in Ref.~\cite{Farry:2019rfg} for the
$6\invfb$ Run 2 dataset recorded by \lhcb.
The effect of these parameter variations on the \ptW distribution is shown in
Fig.~\ref{fig:Wpt_spectrum_Wp}.

These events are reweighted to different values of \mW using a relativistic
Breit-Wigner function with mass-dependent width,
\begin{equation*}
    \left((m^2 - m_{\PW}^{2})^2 + m^4\Gamma_{\PW}^{2}/m_{\PW}^{2}\right)^{-1},
\end{equation*}
where the \PW boson width, $\Gamma_{\PW}$, is fixed to its nominal value and 
$m$ denotes the \PW propagator mass.
Reweighting to arbitrary values of the nuisance parameters \as and \IKT is
based on three-dimensional histograms of the \PW propagator mass\footnote{As reported in the 
\pythia event history.}, rapidity and \pt that have been populated with the events from each
point on the $4\times4$ grid.
These are interpolated to the desired values of \as and \IKT using a two-dimensional cubic
spline.

\section{Fitting method}
\label{sec:fitting}
\begin{figure}
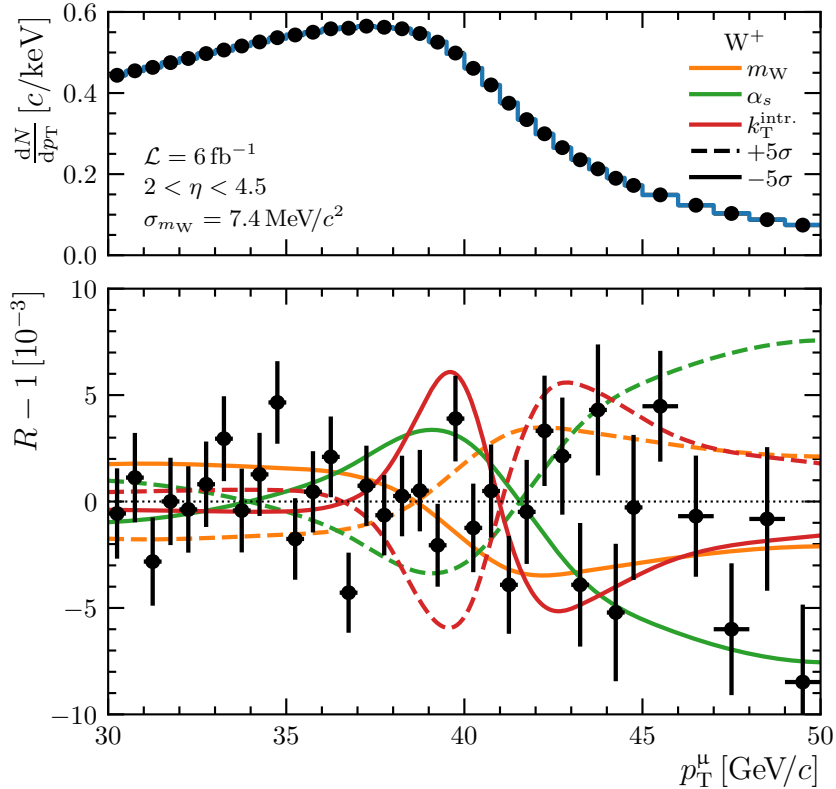

    \centering
    \includegraphics{{{massfit_cl908617.41_Wp}}}
    \caption{Illustrative fit result from a simultaneous fit to the \Wp (shown) and \Wm (see
             Appendix~\ref{app:extra_massfit_plots}) \ptmu distributions.
             This fit assumes the statistics and fiducial region of Ref.~\cite{Farry:2019rfg}.
             The lower panel shows $R-1$, where $R$ denotes the various curves divided by the
             best-fit distribution.
             The coloured curves illustrate the variation in the template distribution when
             the parameter given in the legend is varied by $\pm5\sigma$, where $\sigma$ is
             the uncertainty reported by the fit.}
    \label{fig:Wp_pT_fit_example}
\end{figure}
The values of \mW and the nuisance parameters \as and \IKT are determined using a
binned maximum likelihood fit to \ptmu.
In each fit, the signal shape is described using Monte Carlo template events, which 
are reweighted  on the fly as the values of \mW, \as and \IKT vary.
The Beeston-Barlow ``lite'' prescription~\cite{Barlow:1993dm,Conway:2011in} is used 
to account for the finite Monte Carlo statistics in the signal templates.
An example fit is shown in Fig.~\ref{fig:Wp_pT_fit_example}, where all three of
\mW, \as and \IKT are allowed to vary, and the pseudodata statistics mirror
Ref.~\cite{Farry:2019rfg}.

\newcommand{\pullcaption}[1]{Normalised residuals of #1 using the baseline fit configuration
                             of Sect.~\ref{sec:fitting}.
                            The mean and spread of the unbinned data, and a corresponding 
                            Gaussian curve, are overlaid.}
\begin{figure}
    \centering
    \begin{minipage}[t]{.5\textwidth}
        \centering
        \includegraphics{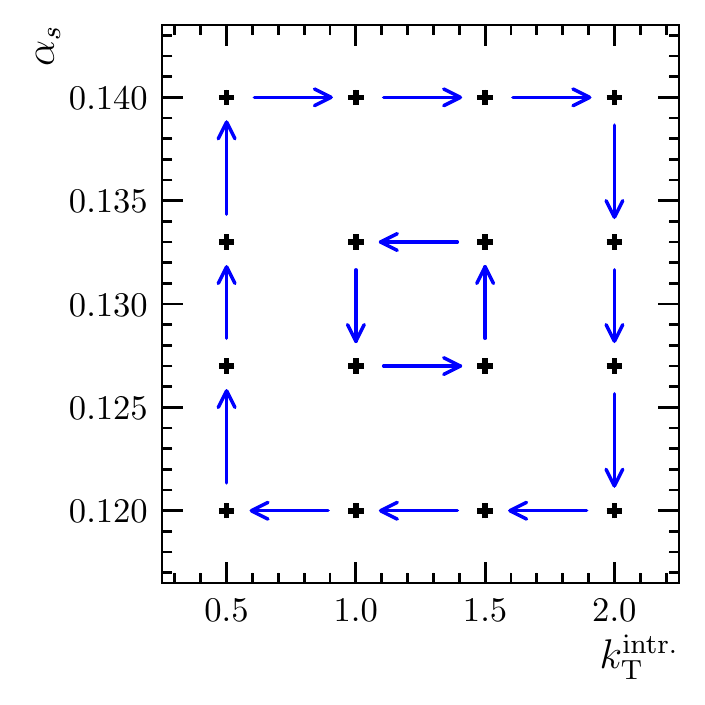}
        \caption{Illustration of relationship between the grid point from which
                 pseudodata is drawn (arrow head) and that from which the signal
                 template events are taken (arrow tail). The choice of nearby points
                 maximises the statistical power of the template events.}
        \label{fig:grid_circle}
    \end{minipage}%
    \begin{minipage}[t]{.5\textwidth}
        \centering
        \includegraphics{{{nuisancetoys_Wpm__13__Pythia__8.235__NNPDF23_lo_as_0130_qed__0__pT_30.0_50.0__float_IKT_as_mW__scale0.25__mW_pulls}}}
        \captionsetup{width=0.8\linewidth}
        \captionof{figure}{\pullcaption{\mW}}
        \label{fig:baseline_mW_pulls}
    \end{minipage}
\end{figure}
The studies in this paper are based on pseudodata fits, where in each fit the
pseudodata are drawn from one point on the $4\times4$ grid, and the signal
templates are based on events from a different point on the grid.
The pairs of grid points are chosen according to the scheme illustrated in 
Fig.~\ref{fig:grid_circle}.
The number of independent pseudodata fits that can be run, therefore, scales
inversely with the desired statistics in each fit.
The baseline configuration scales down the statistics assumed in 
Ref.~\cite{Farry:2019rfg} by a factor of four in order to boost the number of
independent pseudoexperiments that can be run.
The number of signal template events is limited to a maximum of ten times the
pseudodata yield.

\section{Pseudoexperiment results}
The baseline configuration for the results in this paper is to adopt the
$30 < \ptmu < 50\gevc$ and $2 < \eta < 4.5$  kinematic region chosen by
Ref.~\cite{Farry:2019rfg}, with pseudodata statistics reduced by a factor four
with respect to that study as noted in Sect.~\ref{sec:fitting}.
The baseline choice is to allow three physical parameters to float in each fit:
\mW, \as and \IKT.  Various changes to the kinematic region and the choice of free parameters
are also explored.

\begin{figure}
    \centering
    \begin{minipage}[t]{.5\textwidth}
        \centering
        \includegraphics{{{nuisancetoys_Wpm__13__Pythia__8.235__NNPDF23_lo_as_0130_qed__0__pT_30.0_50.0__float_IKT_as_mW__scale0.25__as_pulls}}}
        \captionsetup{width=0.8\linewidth}
        \captionof{figure}{\pullcaption{\as}}
        \label{fig:baseline_as_pulls}
    \end{minipage}%
    \begin{minipage}[t]{.5\textwidth}
        \centering
        \includegraphics{{{nuisancetoys_Wpm__13__Pythia__8.235__NNPDF23_lo_as_0130_qed__0__pT_30.0_50.0__float_IKT_as_mW__scale0.25__IKT_pulls}}}
        \captionsetup{width=0.8\linewidth}
        \captionof{figure}{\pullcaption{\IKT}}
        \label{fig:baseline_IKT_pulls}
    \end{minipage}
\end{figure}

\begin{figure}
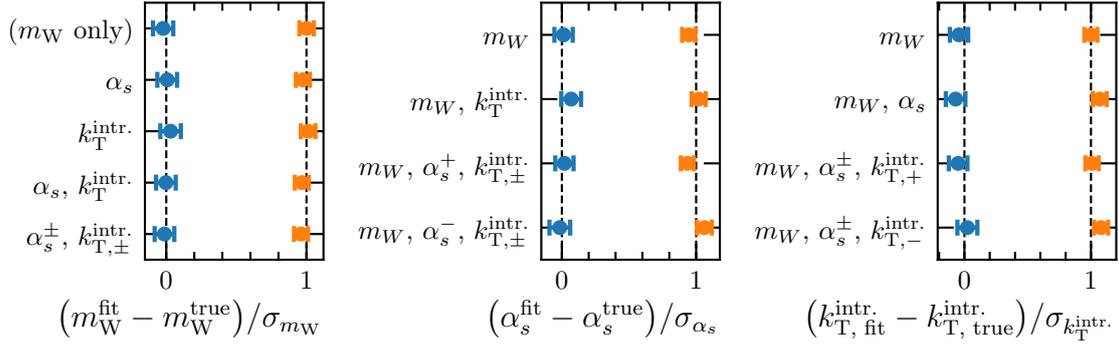

    \centering
    \includegraphics{{{nuisancetoys_Wpm__13__Pythia__8.235__NNPDF23_lo_as_0130_qed__0__pT_30.0_50.0__scale0.25_pull_summary}}}
    \caption{Summary of the mean (blue) and width (orange) of the normalised
             residual distributions obtained from pseudoexperiments with
             different sets of parameters allowed to vary.
             The $y$ axis labels indicate which parameters are free to vary in
             addition to the parameter shown on the $x$ axis.
             Where signs appear in parameter names, such as \asp and \IKTm, this
             indicates that the parameter may take different values for \Wp and 
             \Wm, and \aspm is shorthand for \asp, \asm and so on.
             The baseline configuration corresponds to the fourth row in the
             leftmost figure, and the second row in the centre and right 
             figures.}
    \label{fig:different_floating_pull_summary}
\end{figure}
This baseline configuration produces unbiased results with good statistical 
coverage, as illustrated by Figs.~\ref{fig:baseline_mW_pulls}, 
\ref{fig:baseline_as_pulls} and  \ref{fig:baseline_IKT_pulls}, where results
from every point on the $4\times4$ grid are combined.
With the baseline configuration and available yields there are 192 independent
pseudodatasets, of which 191 survive minimal quality requirements.
The uncertainties are found to be well approximated by symmetric Gaussian 
behaviour.
For brevity, in the rest of the paper, when we consider departures from the baseline 
configuration, such distributions are summarised by their means and widths.
For example, variations in the number of fit parameters are shown in
Fig.~\ref{fig:different_floating_pull_summary}, indicating that the fit procedure
performs well under all considered variations, the most of extreme of which is
to simultaneously fit \mW, and separate values of the nuisance parameters \as
and \IKT for each \PW boson charge.
Further results illustrating the stability of the fit procedure are given in 
Appendix \ref{app:extra_result_plots}.

\begin{figure}
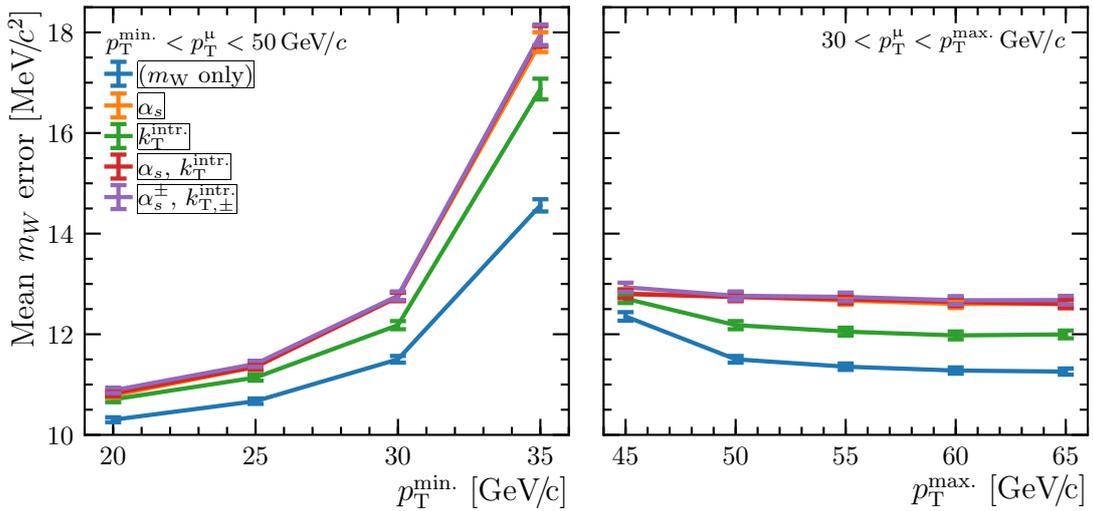

    \centering
    \includegraphics{{{nuisancetoys_Wpm__13__Pythia__8.235__NNPDF23_lo_as_0130_qed__0__pT_30.0_50.0__scale0.25__mW_error}}}
    \caption{Variation of the statistical uncertainty on \mW obtained from     
             several fit configurations, illustrated as a function of the fit
             range in \ptmu.
             As in Fig.~\ref{fig:different_floating_pull_summary}, the legend
             indicates which parameters are free to vary in addition to the
             parameter, \mW, shown on the $y$ axis.
             The baseline configuration corresponds to the red curve at $\ptmin~(\ptmax) = 30~(50)\gevc$ in the left (right) figure.}
    \label{fig:fit_range_mW_error}
\end{figure}
Having demonstrated that the pseudoexperiment setup performs well, it is 
interesting to explore how the fit results depend on choices such as the \ptmu
fit range and the number of freely varying nuisance parameters.
One such study is shown in Fig.~\ref{fig:fit_range_mW_error}, which shows the
the average statistical uncertainty on \mW for several choices of fit range and 
fit parameters.
This shows that the proposed method incurs only a modest degradation in 
statistical precision with respect to the simplest \mW-only fit configuration,
and interestingly that allowing the \Wp and \Wm to each take their own value of
the two nuisance parameters has a negligible effect.

\begin{figure}
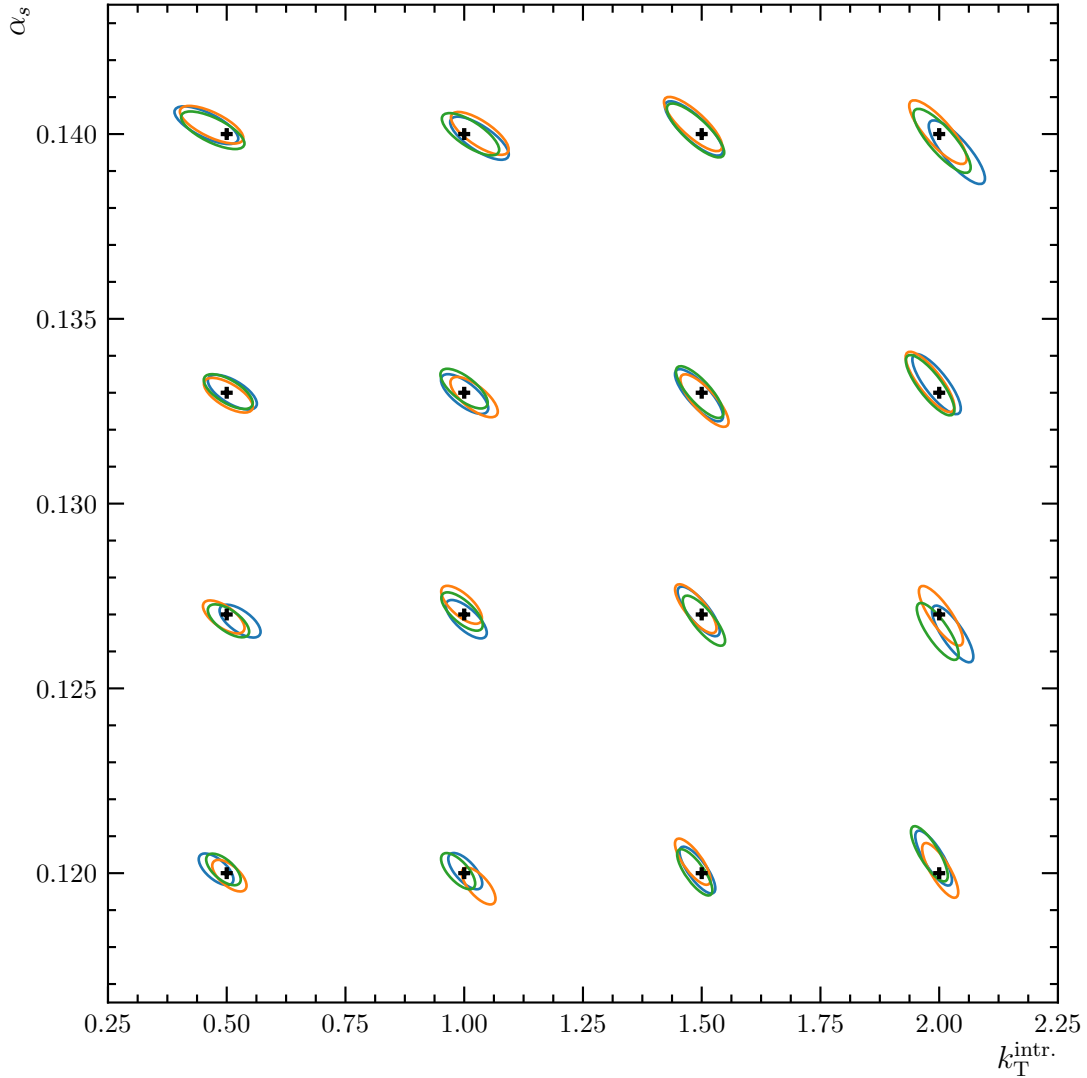

    \centering
    \includegraphics{{{nuisancetoys_scatter__Wpm__13__Pythia__8.235__NNPDF23_lo_as_0130_qed__0__pT_30.0_50.0__float_IKT_as_mW__scale1.00__IKT__as}}}
    \caption{Illustration of the $4\times4$ interpolation grid and the Gaussian
             $3\sigma$ error ellipses obtained from each pseudoexperiment, showing the
             significant anti-correlation between the two nuisance parameters.
             Here the pseudoexperiments correspond to the event yields of 
             Ref.~\cite{Farry:2019rfg}, \ie they are a factor four higher than 
             the baseline configuration.
             The different colours have no particular meaning, and simply serve
             to differentiate the different pseudoexperiment results.}
    \label{fig:scatter_with_ellipses}
\end{figure}
\begin{figure}
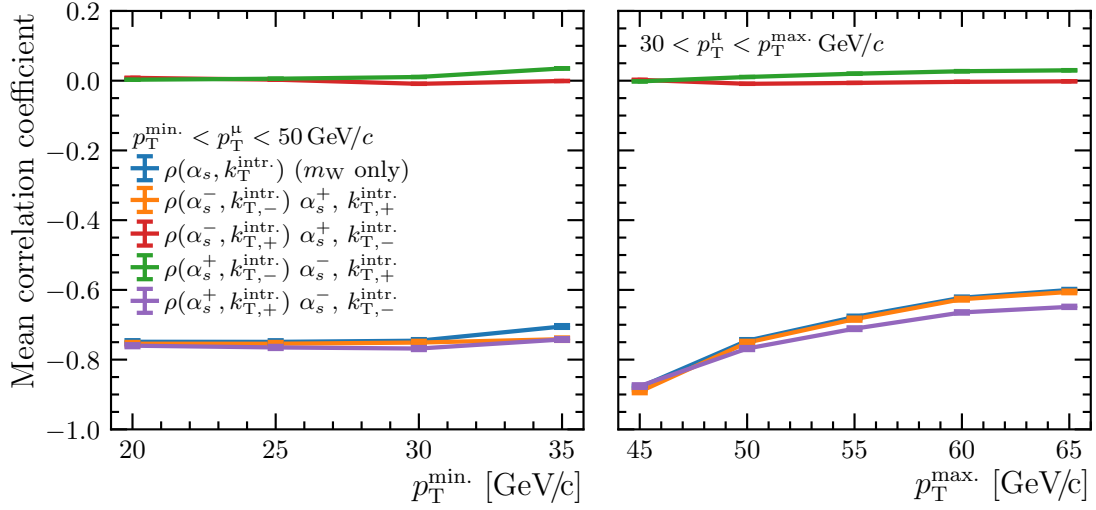

    \centering
    \includegraphics{{{nuisancetoys_Wpm__13__Pythia__8.235__NNPDF23_lo_as_0130_qed__0__pT_30.0_50.0__scale0.25_IKT_as_corr}}}
    \caption{Variation of the \as--\IKT correlation obtained from several fit
             configurations, illustrated as a function of the fit range in 
             \ptmu.
             The meaning of the superscript charges is defined in 
             Fig.~\ref{fig:different_floating_pull_summary} and the legend entries
             are described in Fig.~\ref{fig:fit_range_mW_error}.}
    \label{fig:as_IKT_correlation}
\end{figure}
\begin{figure}
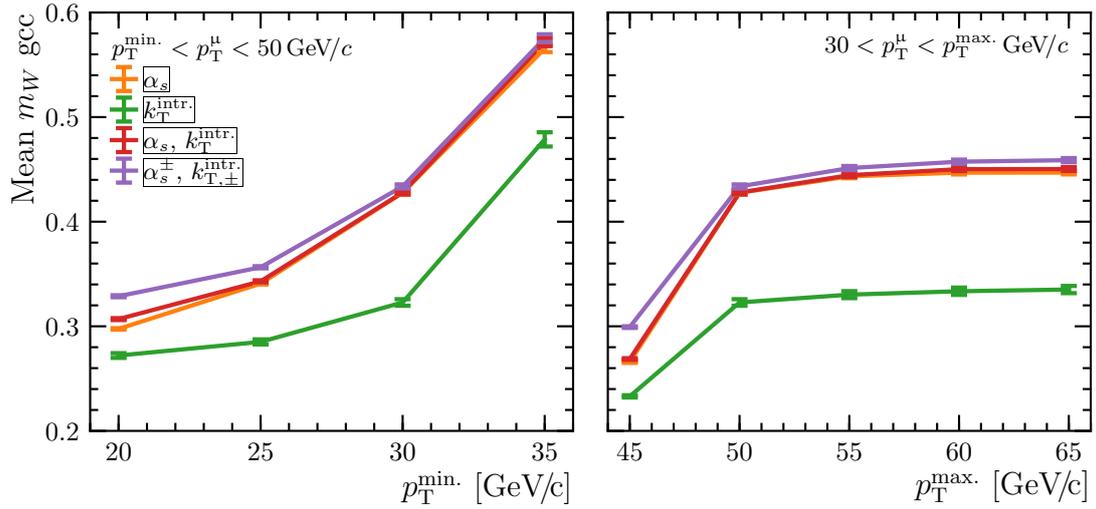

    \centering
    \includegraphics{{{nuisancetoys_Wpm__13__Pythia__8.235__NNPDF23_lo_as_0130_qed__0__pT_30.0_50.0__scale0.25__mW_gcc}}}
    \caption{Variation of the global correlation coefficient (gcc) of \mW
             obtained from several fit configurations, illustrated as a function
             of the fit range in \ptmu.
             The meaning of the superscript charges is defined in 
             Fig.~\ref{fig:different_floating_pull_summary} and the legend entries
             are described in Fig.~\ref{fig:fit_range_mW_error}.
             Note that, as reported in Figs.~\ref{fig:fit_range_as_error} and
             \ref{fig:fit_range_IKT_error}, sensitivity to \as and \IKT deteriorates
             significantly at $\ptmax = 45\gevc$.}
    \label{fig:mW_gcc}
\end{figure}
The two nuisance parameters chosen for this study exhibit a significant
anti-correlation, as might be expected from Fig.~\ref{fig:Wp_pT_fit_example}, 
which is illustrated in Fig.~\ref{fig:scatter_with_ellipses}.
The fit performance distributions already shown indicate that this is not a
major problem, but it can also be seen in Fig.~\ref{fig:as_IKT_correlation} that
increasing the upper \ptmu limit would reduce this correlation, as the large
\ptmu range is principally sensitive to \as.
The extent to which the proposed method can disentangle \mW from the other QCD
nuisance parameters can also be probed by examining the global correlation
coefficient of \mW~\cite{Eadie:100342}.
This is defined as the correlation between \mW and the linear combination of all
other fit parameters that it is most strongly correlated with, and it is shown
for \mW in Fig.~\ref{fig:mW_gcc}.
It can be seen that reducing \ptmin tends to reduce the degeneracy of the fit
parameters.

It is also confirmed that adopting the event yields of Ref.~\cite{Farry:2019rfg}, \ie 
increasing those of the baseline configuration by a factor four, does not introduce any bias 
or coverage problems, and it is this higher-statistics configuration that is illustrated in 
Figs.~\ref{fig:Wp_pT_fit_example} and \ref{fig:Wm_pT_fit_example}.

Several additional figures showing the various parameter uncertainties, their
correlations and the variation of these quantities with different fit 
configurations are included in Appendix~\ref{app:extra_result_plots}.

\section{Conclusions}
We have demonstrated that it is possible to simultaneously determine both the \PW boson mass, \mW,
and nuisance parameters relating to its \pt spectrum using a fit to the \ptmu spectrum with only a 
small inflation of the statistical uncertainty on \mW.
We find that, for the specific parameters that were chosen to illustrate the technique, the
simultaneous fit is well-behaved and that for most reasonable choices of the \ptmu fit range the
fits have little trouble disentangling variations in \mW from those in the \ptW model.
The study considers variations of the nuisance parameters that correspond to variations in the \ptW 
spectrum that are large compared to the uncertainty of state-of-the-art predictions, indicating that
the proposed technique is sufficiently powerful to enable a precise measurement of \mW.

In an actual measurement of \mW it would, of course, be preferable to apply the same technique using
predictions from tools that contain higher order electroweak and QCD corrections, which naturally
leads to the question of what parameters can legitimately be varied in this case.
The examples that have been shown to work well with \pythia in this study could provide a useful
template: even in the more accurate calculations it should be possible to identify a \IKT-like
nonperturbative smearing, and to vary the strong coupling constant, but other choices may prove to
be optimal for different tools.
A larger number of nuisance parameters could also be varied simultaneously, if this was 
well motivated for a particular tool; the implementation used in this paper in theory has no upper
limit, but in practice it is limited to varying a maximum of 3--4 parameters in addition to \mW.

It will also be interesting to explore how this method can be combined with the techniques explored in
Ref.~\cite{Farry:2019rfg} for reducing PDF uncertainties using in-situ constraints, and it is
important to verify that the inclusion of realistic levels of QCD and electroweak backgrounds does
not adversely affect the performance of the method.

In summary, the proposed technique performs well using pseudodata generated with \pythia, and 
appears to provide a possible route to a precise measurement of \mW that is less reliant on accurate
modelling of the differences between \PW and \Zz boson production.

\section*{Acknowledgements}
\noindent We thank W. Barter, M. Charles, S. Farry, R. Hunter, M. Pili, F. Tackmann
and A. Vicini for their helpful comments and suggestions during the preparation of
this manuscript.
OL thanks the CERN LBD group for their support during the initial stages of this 
work, and MV thanks the Science and Technologies Facilities Council for their support
through an Ernest Rutherford Fellowship.

\clearpage
\section*{Appendices}
\appendix
Appendix~\ref{app:pythia_params} provides additional detail regarding the \pythia
configuration used throughout this study, while 
Appendix~\ref{app:extra_massfit_plots} includes the \Wm counterpart of the
illustrative \Wp fit distribution shown in the main text.
Finally, Appendix~\ref{app:extra_result_plots} includes additional results from
the ensemble of pseudoexperiments.
\section{\pythia tuning parameters}
\label{app:pythia_params}
The quantity \as used throughout this paper refers to the \pythia configuration options
\texttt{TimeShower:alphaSvalue} and \texttt{SpaceShower:alphaSvalue}, while the 
quantity \IKT is a scale factor applied to the configuration options
\begin{align*}
    \texttt{BeamRemnants:halfScaleForKT} &= 1.5\times\IKT, \\
    \texttt{BeamRemnants:primordialKTsoft} &= 0.9\times\IKT, \\
    \texttt{BeamRemnants:primordialKThard} &= 1.8\times\IKT. \\
\end{align*}
The $4\times4$ grid consists of $\as \in \{0.120, 0.127, 0.133, 0.140\}$ and
$\IKT \in \{0.5, 1.0, 1.5, 2.0\}$.
With the exception of these parameters, the default tuning of \pythia 8.235 is used.
\section{Additional kinematic distributions}
\label{app:extra_massfit_plots}
\begin{figure}
    \centering
    \includegraphics{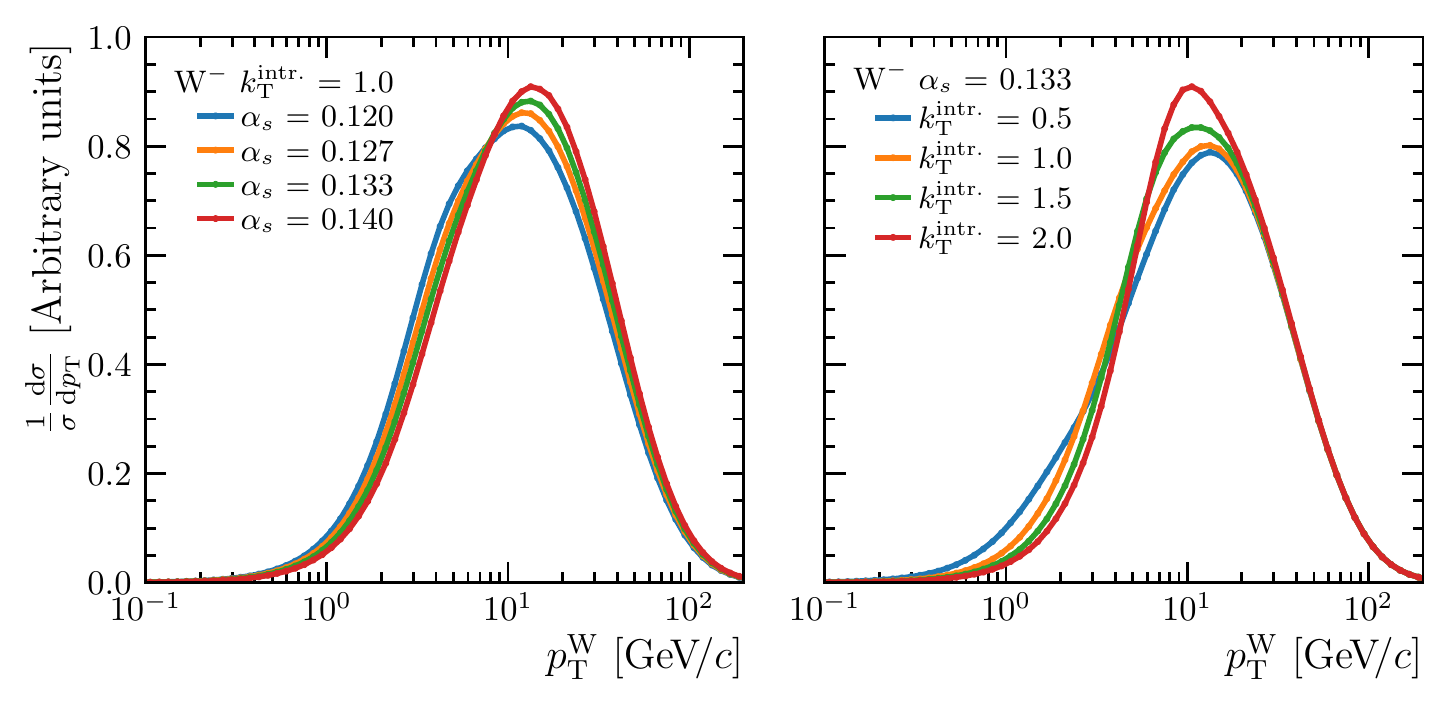}
    \caption{Illustration of variations in the \Wm boson \pt spectrum corresponding
             to variations in the \as (left) and \IKT (right) nuisance parameters.
             No kinematic requirements have been placed on the \Wm decay products.
             This is the \Wm analogue of the \Wp distributions shown in Fig.~\ref{fig:Wpt_spectrum_Wp}.}
    \label{fig:Wpt_spectrum_Wm}
\end{figure}

\begin{figure}
    \centering
    \includegraphics{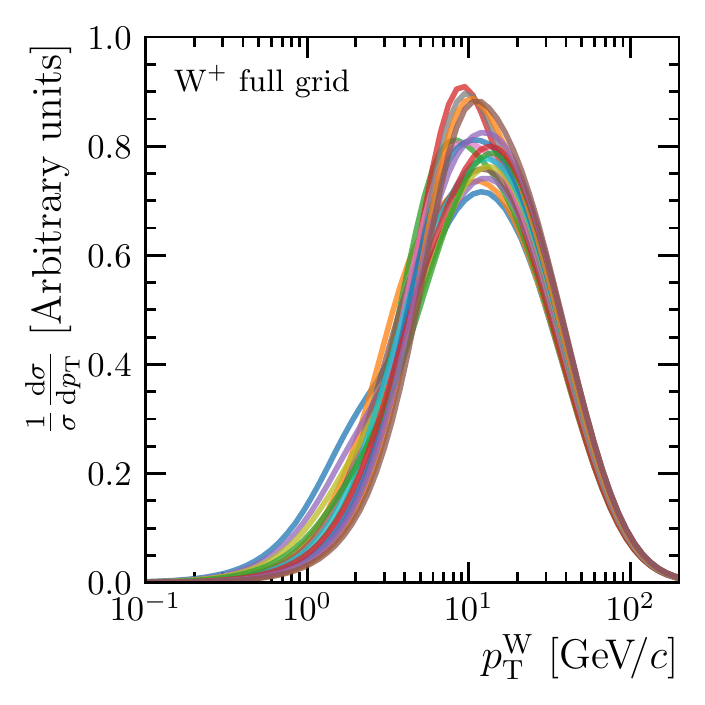}
    \caption{Illustration of variations in the \Wp boson \pt spectrum corresponding
             to all 16 points on the $4\times 4$ grid of \as and \IKT values.}
    \label{fig:Wpt_spectrum_Wp_all}
\end{figure}

\begin{figure}
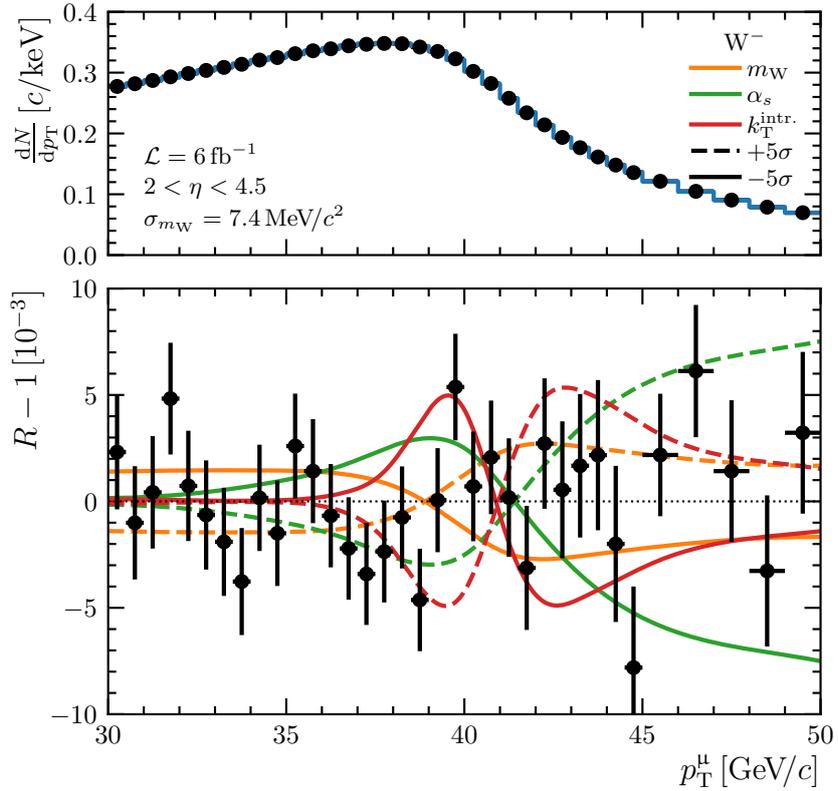

    \centering
    \includegraphics{{{massfit_cl908617.41_Wm}}}
    \caption{Illustrative fit result from a simultaneous fit to the \Wp 
             (see Sect.~\ref{sec:fitting}) and \Wm (shown) \ptmu distributions.
             Further information is given alongside 
             Fig.~\ref{fig:Wp_pT_fit_example}.}
    \label{fig:Wm_pT_fit_example}
\end{figure}

This section contains Fig.~\ref{fig:Wpt_spectrum_Wm}, which is the \Wm counterpart
to the \Wp boson \pt distributions shown in Fig.~\ref{fig:Wpt_spectrum_Wp}, 
Fig.~\ref{fig:Wpt_spectrum_Wp_all}, which is a more verbose analogue to 
Fig.~\ref{fig:Wpt_spectrum_Wp}, and Fig.~\ref{fig:Wm_pT_fit_example}, which is 
the \Wm counterpart of the \Wp fit, Fig.~\ref{fig:Wp_pT_fit_example}, shown in the 
main body of the paper.
\section{Additional pseudoexperiment results}
\label{app:extra_result_plots}
\begin{figure}
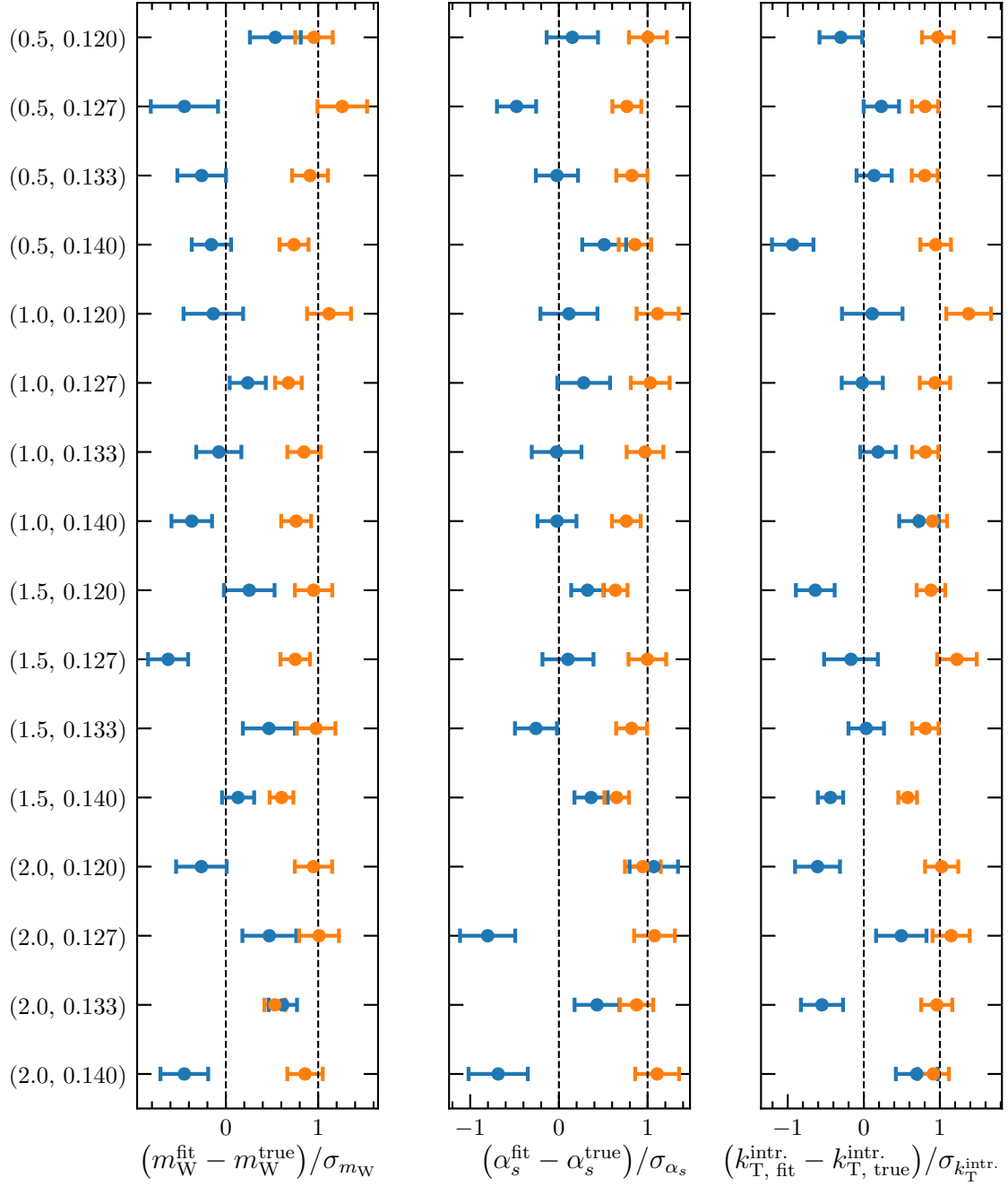

    \centering
    \includegraphics{{{nuisancetoys_Wpm__13__Pythia__8.235__NNPDF23_lo_as_0130_qed__0__pT_30.0_50.0__float_IKT_as_mW__scale0.25_pull_summary}}}
    \caption{Summary of the mean (blue) and width (orange) of the normalised
             residual distributions shown in Figs.~\ref{fig:baseline_mW_pulls},
             \ref{fig:baseline_as_pulls} and \ref{fig:baseline_IKT_pulls} as a
             function of position on the $4\times4$ grid.
             The $y$ axis labels give the $(\IKT, \as)$ coordinates from which
             the pseudodata are drawn.}
    \label{fig:pull_summary_baseline_split_by_grid}
\end{figure}

\begin{figure}
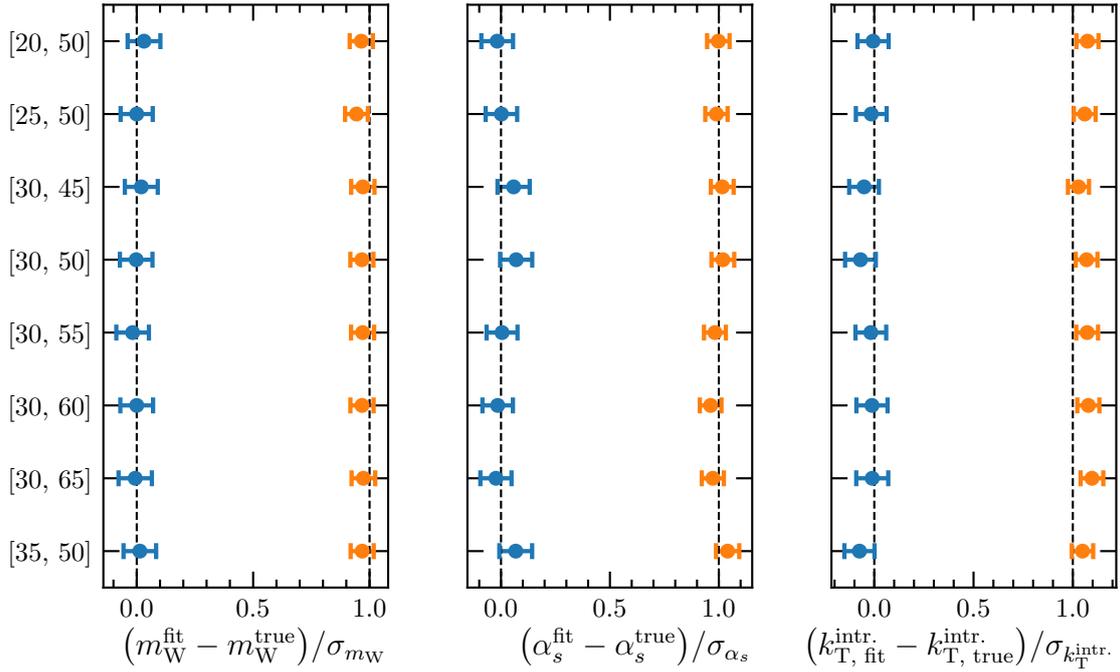

    \centering
    \includegraphics{{{nuisancetoys_Wpm__13__Pythia__8.235__NNPDF23_lo_as_0130_qed__0__float_IKT_as_mW__scale0.25_pull_summary}}}
    \caption{Summary of the mean (blue) and width (orange) of the normalised
             residual distributions obtained from pseudoexperiments with 
             different \ptmu fit range choices, which are given by the $y$ axis
             labels.
             The row labelled $[30,50]$ corresponds to the distributions shown
             in Figs.~\ref{fig:baseline_mW_pulls}, \ref{fig:baseline_as_pulls}
             and \ref{fig:baseline_IKT_pulls}.}
    \label{fig:pull_summary_baseline_split_by_pT}
\end{figure}

\begin{figure}
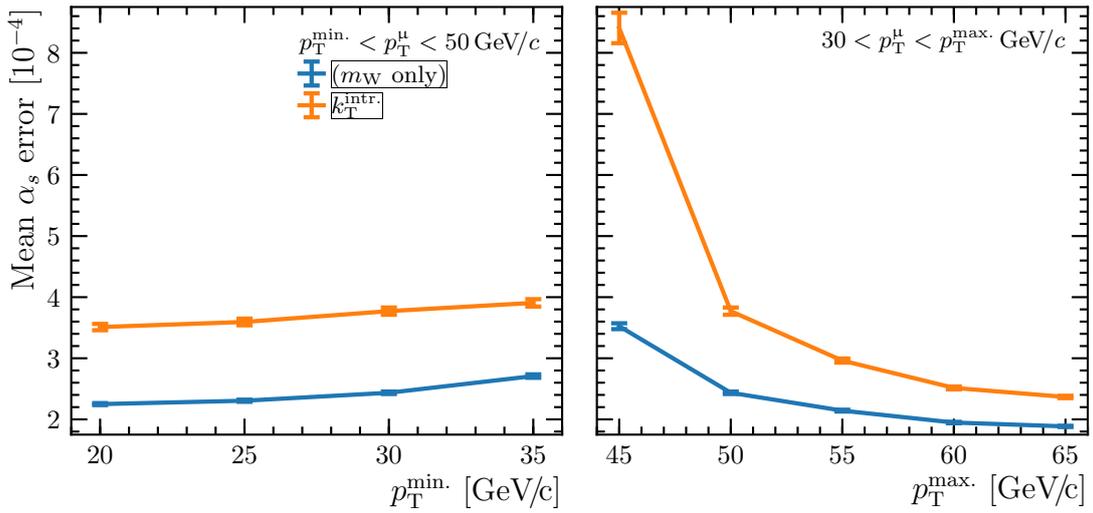

    \centering
    \includegraphics{{{nuisancetoys_Wpm__13__Pythia__8.235__NNPDF23_lo_as_0130_qed__0__pT_30.0_50.0__scale0.25__as_error}}}
    \caption{Variation of the fitted uncertainty on \as with fit configuration and \ptmu range.
             This is the analogue of Fig.~\ref{fig:fit_range_mW_error} for \as.}
    \label{fig:fit_range_as_error}
\end{figure}

This section contains additional results showing the stability of the fit 
procedure with respect to different departures from the baseline configuration.
Figure~\ref{fig:pull_summary_baseline_split_by_grid} shows that the fit procedure
is reasonably stable as the absolute values of the nuisance parameters vary across
the $4\times4$ grid of \as and \IKT values.
Figure~\ref{fig:pull_summary_baseline_split_by_pT} shows the fit procedure
remains stable when the allowed range of the muon transverse momentum, \ptmu, is
varied.
Figures~\ref{fig:fit_range_as_error} and \ref{fig:fit_range_IKT_error} show the dependence of the
uncertainties with which \as and \IKT are determined on the fit configuration, while 
Figs.~\ref{fig:fit_range_as_gcc} and \ref{fig:fit_range_IKT_gcc} show the global correlation
coefficients of these parameters.
Finally, Figs.~\ref{fig:mW_as_correlation} and \ref{fig:mW_IKT_correlation} show the correlation
coefficients between \mW and \as and \IKT.
\begin{figure}
    \centering
    \includegraphics{{{nuisancetoys_Wpm__13__Pythia__8.235__NNPDF23_lo_as_0130_qed__0__pT_30.0_50.0__scale0.25__IKT_error}}}
    \caption{Variation of the fitted uncertainty on \IKT with fit configuration and \ptmu range.
             This is the analogue of Fig.~\ref{fig:fit_range_mW_error} for \IKT.}
    \label{fig:fit_range_IKT_error}
\end{figure}
\begin{figure}
    \centering
    \includegraphics{{{nuisancetoys_Wpm__13__Pythia__8.235__NNPDF23_lo_as_0130_qed__0__pT_30.0_50.0__scale0.25__as_gcc}}}
    \caption{Variation of the global correlation coefficient (gcc) of \as with fit configuration
             and \ptmu range. This is the analogue of Fig.~\ref{fig:mW_gcc} for \as.}
    \label{fig:fit_range_as_gcc}
\end{figure}
\begin{figure}
    \centering
    \includegraphics{{{nuisancetoys_Wpm__13__Pythia__8.235__NNPDF23_lo_as_0130_qed__0__pT_30.0_50.0__scale0.25__IKT_gcc}}}
    \caption{Variation of the global correlation coefficient (gcc) of \IKT with fit configuration
             and \ptmu range. This is the analogue of Fig.~\ref{fig:mW_gcc} for \IKT.}
    \label{fig:fit_range_IKT_gcc}
\end{figure}
\begin{figure}
    \centering
    \includegraphics{{{nuisancetoys_Wpm__13__Pythia__8.235__NNPDF23_lo_as_0130_qed__0__pT_30.0_50.0__scale0.25_as_mW_corr}}}
    \caption{Variation of the \mW--\as correlation obtained from several fit configurations, 
             illustrated as a function of the fit range in \ptmu.
             The meaning of the superscript charges is defined in 
             Fig.~\ref{fig:different_floating_pull_summary}.
             This is the \mW--\as analogue of Fig.~\ref{fig:as_IKT_correlation}.}
    \label{fig:mW_as_correlation}
\end{figure}
\begin{figure}
    \centering
    \includegraphics{{{nuisancetoys_Wpm__13__Pythia__8.235__NNPDF23_lo_as_0130_qed__0__pT_30.0_50.0__scale0.25_IKT_mW_corr}}}
    \caption{Variation of the \mW--\IKT correlation obtained from several fit configurations, 
             illustrated as a function of the fit range in \ptmu.
             The meaning of the superscript charges is defined in 
             Fig.~\ref{fig:different_floating_pull_summary}.
             This is the \mW--\IKT analogue of Fig.~\ref{fig:as_IKT_correlation}.}
    \label{fig:mW_IKT_correlation}
\end{figure}

\addcontentsline{toc}{section}{References}
\bibliographystyle{LHCb}
\bibliography{main}

\ifx\mcitethebibliography\mciteundefinedmacro
\PackageError{LHCb.bst}{mciteplus.sty has not been loaded}
{This bibstyle requires the use of the mciteplus package.}\fi
\providecommand{\href}[2]{#2}
\begin{mcitethebibliography}{10}
\mciteSetBstSublistMode{n}
\mciteSetBstMaxWidthForm{subitem}{\alph{mcitesubitemcount})}
\mciteSetBstSublistLabelBeginEnd{\mcitemaxwidthsubitemform\space}
{\relax}{\relax}

\bibitem{Haller:2018nnx}
J.~Haller {\em et~al.}, \ifthenelse{\boolean{articletitles}}{\emph{{Update of
  the global electroweak fit and constraints on two-Higgs-doublet models}},
  }{}\href{https://doi.org/10.1140/epjc/s10052-018-6131-3}{Eur.\ Phys.\ J.\
  \textbf{C78} (2018) 675},
  \href{http://arxiv.org/abs/1803.01853}{{\normalfont\ttfamily
  arXiv:1803.01853}}\relax
\mciteBstWouldAddEndPuncttrue
\mciteSetBstMidEndSepPunct{\mcitedefaultmidpunct}
{\mcitedefaultendpunct}{\mcitedefaultseppunct}\relax
\EndOfBibitem
\bibitem{Aaltonen:2012bp}
CDF collaboration, T.~Aaltonen {\em et~al.},
  \ifthenelse{\boolean{articletitles}}{\emph{{Precise measurement of the $W$
  boson mass with the CDF II detector}},
  }{}\href{https://doi.org/10.1103/PhysRevLett.108.151803}{Phys.\ Rev.\ Lett.\
  \textbf{108} (2012) 151803},
  \href{http://arxiv.org/abs/1203.0275}{{\normalfont\ttfamily
  arXiv:1203.0275}}\relax
\mciteBstWouldAddEndPuncttrue
\mciteSetBstMidEndSepPunct{\mcitedefaultmidpunct}
{\mcitedefaultendpunct}{\mcitedefaultseppunct}\relax
\EndOfBibitem
\bibitem{Abazov:2012bv}
D0 collaboration, V.~M. Abazov {\em et~al.},
  \ifthenelse{\boolean{articletitles}}{\emph{{Measurement of the W boson mass
  with the D0 detector}},
  }{}\href{https://doi.org/10.1103/PhysRevLett.108.151804}{Phys.\ Rev.\ Lett.\
  \textbf{108} (2012) 151804},
  \href{http://arxiv.org/abs/1203.0293}{{\normalfont\ttfamily
  arXiv:1203.0293}}\relax
\mciteBstWouldAddEndPuncttrue
\mciteSetBstMidEndSepPunct{\mcitedefaultmidpunct}
{\mcitedefaultendpunct}{\mcitedefaultseppunct}\relax
\EndOfBibitem
\bibitem{Aaboud:2017svj}
ATLAS collaboration, M.~Aaboud {\em et~al.},
  \ifthenelse{\boolean{articletitles}}{\emph{{Measurement of the W boson mass
  in pp collisions at $\sqrt{s}$ = 7 Te\!V with the ATLAS detector}},
  }{}\href{https://doi.org/10.1140/epjc/s10052-018-6354-3,
  10.1140/epjc/s10052-017-5475-4}{Eur.\ Phys.\ J.\  \textbf{C78} (2018) 110},
  \href{http://arxiv.org/abs/1701.07240}{{\normalfont\ttfamily
  arXiv:1701.07240}}, [Erratum: Eur. Phys. J.C78,no.11,898(2018)]\relax
\mciteBstWouldAddEndPuncttrue
\mciteSetBstMidEndSepPunct{\mcitedefaultmidpunct}
{\mcitedefaultendpunct}{\mcitedefaultseppunct}\relax
\EndOfBibitem
\bibitem{PDG2018}
Particle Data Group, M.~Tanabashi {\em et~al.},
  \ifthenelse{\boolean{articletitles}}{\emph{{\href{http://pdg.lbl.gov/}{Review
  of particle physics}}},
  }{}\href{https://doi.org/10.1103/PhysRevD.98.030001}{Phys.\ Rev.\
  \textbf{D98} (2018) 030001}\relax
\mciteBstWouldAddEndPuncttrue
\mciteSetBstMidEndSepPunct{\mcitedefaultmidpunct}
{\mcitedefaultendpunct}{\mcitedefaultseppunct}\relax
\EndOfBibitem
\bibitem{Bozzi:2015zja}
G.~Bozzi, L.~Citelli, M.~Vesterinen, and A.~Vicini,
  \ifthenelse{\boolean{articletitles}}{\emph{{Prospects for improving the LHC W
  boson mass measurement with forward muons}},
  }{}\href{https://doi.org/10.1140/epjc/s10052-015-3810-1}{Eur.\ Phys.\ J.\
  \textbf{C75} (2015) 601},
  \href{http://arxiv.org/abs/1508.06954}{{\normalfont\ttfamily
  arXiv:1508.06954}}\relax
\mciteBstWouldAddEndPuncttrue
\mciteSetBstMidEndSepPunct{\mcitedefaultmidpunct}
{\mcitedefaultendpunct}{\mcitedefaultseppunct}\relax
\EndOfBibitem
\bibitem{Farry:2019rfg}
S.~Farry, O.~Lupton, M.~Pili, and M.~Vesterinen,
  \ifthenelse{\boolean{articletitles}}{\emph{{Understanding and constraining
  the PDF uncertainties in a $W$ boson mass measurement with forward muons at
  the LHC}}, }{}\href{https://doi.org/10.1140/epjc/s10052-019-6997-8}{Eur.\
  Phys.\ J.\  \textbf{C79} (2019) 497},
  \href{http://arxiv.org/abs/1902.04323}{{\normalfont\ttfamily
  arXiv:1902.04323}}\relax
\mciteBstWouldAddEndPuncttrue
\mciteSetBstMidEndSepPunct{\mcitedefaultmidpunct}
{\mcitedefaultendpunct}{\mcitedefaultseppunct}\relax
\EndOfBibitem
\bibitem{Hamberg:1990np}
R.~Hamberg, W.~L. van Neerven, and T.~Matsuura,
  \ifthenelse{\boolean{articletitles}}{\emph{{A complete calculation of the
  order $\alpha_{s}^{2}$ correction to the Drell-Yan $K$ factor}},
  }{}\href{https://doi.org/10.1016/S0550-3213(02)00814-3,
  10.1016/0550-3213(91)90064-5}{Nucl.\ Phys.\  \textbf{B359} (1991) 343},
  [Erratum: Nucl. Phys.B644,403(2002)]\relax
\mciteBstWouldAddEndPuncttrue
\mciteSetBstMidEndSepPunct{\mcitedefaultmidpunct}
{\mcitedefaultendpunct}{\mcitedefaultseppunct}\relax
\EndOfBibitem
\bibitem{vanNeerven:1991gh}
W.~L. van Neerven and E.~B. Zijlstra,
  \ifthenelse{\boolean{articletitles}}{\emph{{The
  $\mathcal{O}\left(\alpha_{s}^{2}\right)$ corrected Drell-Yan $K$ factor in
  the DIS and MS scheme}},
  }{}\href{https://doi.org/10.1016/j.nuclphysb.2003.12.019,
  10.1016/0550-3213(92)90078-P}{Nucl.\ Phys.\  \textbf{B382} (1992) 11},
  [Erratum: Nucl. Phys.B680,513(2004)]\relax
\mciteBstWouldAddEndPuncttrue
\mciteSetBstMidEndSepPunct{\mcitedefaultmidpunct}
{\mcitedefaultendpunct}{\mcitedefaultseppunct}\relax
\EndOfBibitem
\bibitem{Catani:2009sm}
S.~Catani {\em et~al.}, \ifthenelse{\boolean{articletitles}}{\emph{{Vector
  boson production at hadron colliders: a fully exclusive QCD calculation at
  NNLO}}, }{}\href{https://doi.org/10.1103/PhysRevLett.103.082001}{Phys.\ Rev.\
  Lett.\  \textbf{103} (2009) 082001},
  \href{http://arxiv.org/abs/0903.2120}{{\normalfont\ttfamily
  arXiv:0903.2120}}\relax
\mciteBstWouldAddEndPuncttrue
\mciteSetBstMidEndSepPunct{\mcitedefaultmidpunct}
{\mcitedefaultendpunct}{\mcitedefaultseppunct}\relax
\EndOfBibitem
\bibitem{Gavin:2010az}
R.~Gavin, Y.~Li, F.~Petriello, and S.~Quackenbush,
  \ifthenelse{\boolean{articletitles}}{\emph{{FEWZ 2.0: a code for hadronic Z
  production at next-to-next-to-leading order}},
  }{}\href{https://doi.org/10.1016/j.cpc.2011.06.008}{Comput.\ Phys.\ Commun.\
  \textbf{182} (2011) 2388},
  \href{http://arxiv.org/abs/1011.3540}{{\normalfont\ttfamily
  arXiv:1011.3540}}\relax
\mciteBstWouldAddEndPuncttrue
\mciteSetBstMidEndSepPunct{\mcitedefaultmidpunct}
{\mcitedefaultendpunct}{\mcitedefaultseppunct}\relax
\EndOfBibitem
\bibitem{Anastasiou:2003ds}
C.~Anastasiou, L.~J. Dixon, K.~Melnikov, and F.~Petriello,
  \ifthenelse{\boolean{articletitles}}{\emph{{High precision QCD at hadron
  colliders: electroweak gauge boson rapidity distributions at NNLO}},
  }{}\href{https://doi.org/10.1103/PhysRevD.69.094008}{Phys.\ Rev.\
  \textbf{D69} (2004) 094008},
  \href{http://arxiv.org/abs/hep-ph/0312266}{{\normalfont\ttfamily
  arXiv:hep-ph/0312266}}\relax
\mciteBstWouldAddEndPuncttrue
\mciteSetBstMidEndSepPunct{\mcitedefaultmidpunct}
{\mcitedefaultendpunct}{\mcitedefaultseppunct}\relax
\EndOfBibitem
\bibitem{Boughezal:2015dva}
R.~Boughezal, C.~Focke, X.~Liu, and F.~Petriello,
  \ifthenelse{\boolean{articletitles}}{\emph{{$W$ boson production in
  association with a jet at next-to-next-to-leading order in perturbative
  QCD}}, }{}\href{https://doi.org/10.1103/PhysRevLett.115.062002}{Phys.\ Rev.\
  Lett.\  \textbf{115} (2015) 062002},
  \href{http://arxiv.org/abs/1504.02131}{{\normalfont\ttfamily
  arXiv:1504.02131}}\relax
\mciteBstWouldAddEndPuncttrue
\mciteSetBstMidEndSepPunct{\mcitedefaultmidpunct}
{\mcitedefaultendpunct}{\mcitedefaultseppunct}\relax
\EndOfBibitem
\bibitem{Gehrmann-DeRidder:2017mvr}
A.~Gehrmann-De~Ridder {\em et~al.},
  \ifthenelse{\boolean{articletitles}}{\emph{{Next-to-next-to-leading-order QCD
  corrections to the transverse momentum distribution of weak gauge bosons}},
  }{}\href{https://doi.org/10.1103/PhysRevLett.120.122001}{Phys.\ Rev.\ Lett.\
  \textbf{120} (2018) 122001},
  \href{http://arxiv.org/abs/1712.07543}{{\normalfont\ttfamily
  arXiv:1712.07543}}\relax
\mciteBstWouldAddEndPuncttrue
\mciteSetBstMidEndSepPunct{\mcitedefaultmidpunct}
{\mcitedefaultendpunct}{\mcitedefaultseppunct}\relax
\EndOfBibitem
\bibitem{Dittmaier:2001ay}
S.~Dittmaier and M.~Kramer,
  \ifthenelse{\boolean{articletitles}}{\emph{{Electroweak radiative corrections
  to W boson production at hadron colliders}},
  }{}\href{https://doi.org/10.1103/PhysRevD.65.073007}{Phys.\ Rev.\
  \textbf{D65} (2002) 073007},
  \href{http://arxiv.org/abs/hep-ph/0109062}{{\normalfont\ttfamily
  arXiv:hep-ph/0109062}}\relax
\mciteBstWouldAddEndPuncttrue
\mciteSetBstMidEndSepPunct{\mcitedefaultmidpunct}
{\mcitedefaultendpunct}{\mcitedefaultseppunct}\relax
\EndOfBibitem
\bibitem{Arbuzov:2005dd}
A.~Arbuzov {\em et~al.}, \ifthenelse{\boolean{articletitles}}{\emph{{One-loop
  corrections to the Drell-Yan process in SANC. I. The Charged current case}},
  }{}\href{https://doi.org/10.1140/epjc/s2006-02505-y,
  10.1140/epjc/s10052-007-0225-7}{Eur.\ Phys.\ J.\  \textbf{C46} (2006) 407},
  \href{http://arxiv.org/abs/hep-ph/0506110}{{\normalfont\ttfamily
  arXiv:hep-ph/0506110}}\relax
\mciteBstWouldAddEndPuncttrue
\mciteSetBstMidEndSepPunct{\mcitedefaultmidpunct}
{\mcitedefaultendpunct}{\mcitedefaultseppunct}\relax
\EndOfBibitem
\bibitem{CarloniCalame:2006zq}
C.~M. Carloni~Calame, G.~Montagna, O.~Nicrosini, and A.~Vicini,
  \ifthenelse{\boolean{articletitles}}{\emph{{Precision electroweak calculation
  of the charged current Drell-Yan process}},
  }{}\href{https://doi.org/10.1088/1126-6708/2006/12/016}{JHEP \textbf{0612}
  (2006) 016}, \href{http://arxiv.org/abs/hep-ph/0609170}{{\normalfont\ttfamily
  arXiv:hep-ph/0609170}}\relax
\mciteBstWouldAddEndPuncttrue
\mciteSetBstMidEndSepPunct{\mcitedefaultmidpunct}
{\mcitedefaultendpunct}{\mcitedefaultseppunct}\relax
\EndOfBibitem
\bibitem{Baur:2004ig}
U.~Baur and D.~Wackeroth,
  \ifthenelse{\boolean{articletitles}}{\emph{{Electroweak radiative corrections
  to $p \bar{p} \to W^\pm \to \ell^\pm \nu$ beyond the pole approximation}},
  }{}\href{https://doi.org/10.1103/PhysRevD.70.073015}{Phys.\ Rev.\
  \textbf{D70} (2004) 073015},
  \href{http://arxiv.org/abs/hep-ph/0405191}{{\normalfont\ttfamily
  arXiv:hep-ph/0405191}}\relax
\mciteBstWouldAddEndPuncttrue
\mciteSetBstMidEndSepPunct{\mcitedefaultmidpunct}
{\mcitedefaultendpunct}{\mcitedefaultseppunct}\relax
\EndOfBibitem
\bibitem{Barze:2012tt}
L.~Barze {\em et~al.},
  \ifthenelse{\boolean{articletitles}}{\emph{{Implementation of electroweak
  corrections in the POWHEG BOX: single W production}},
  }{}\href{https://doi.org/10.1007/JHEP04(2012)037}{JHEP \textbf{04} (2012)
  037}, \href{http://arxiv.org/abs/1202.0465}{{\normalfont\ttfamily
  arXiv:1202.0465}}\relax
\mciteBstWouldAddEndPuncttrue
\mciteSetBstMidEndSepPunct{\mcitedefaultmidpunct}
{\mcitedefaultendpunct}{\mcitedefaultseppunct}\relax
\EndOfBibitem
\bibitem{Becher:2011xn}
T.~Becher, M.~Neubert, and D.~Wilhelm,
  \ifthenelse{\boolean{articletitles}}{\emph{{Electroweak Gauge-Boson
  Production at Small $q_T$: Infrared Safety from the Collinear Anomaly}},
  }{}\href{https://doi.org/10.1007/JHEP02(2012)124}{JHEP \textbf{02} (2012)
  124}, \href{http://arxiv.org/abs/1109.6027}{{\normalfont\ttfamily
  arXiv:1109.6027}}\relax
\mciteBstWouldAddEndPuncttrue
\mciteSetBstMidEndSepPunct{\mcitedefaultmidpunct}
{\mcitedefaultendpunct}{\mcitedefaultseppunct}\relax
\EndOfBibitem
\bibitem{Bozzi:2010xn}
G.~Bozzi {\em et~al.}, \ifthenelse{\boolean{articletitles}}{\emph{{Production
  of Drell-Yan lepton pairs in hadron collisions: transverse-momentum
  resummation at next-to-next-to-leading logarithmic accuracy}},
  }{}\href{https://doi.org/10.1016/j.physletb.2010.12.024}{Phys.\ Lett.\
  \textbf{B696} (2011) 207},
  \href{http://arxiv.org/abs/1007.2351}{{\normalfont\ttfamily
  arXiv:1007.2351}}\relax
\mciteBstWouldAddEndPuncttrue
\mciteSetBstMidEndSepPunct{\mcitedefaultmidpunct}
{\mcitedefaultendpunct}{\mcitedefaultseppunct}\relax
\EndOfBibitem
\bibitem{Banfi:2012du}
A.~Banfi, M.~Dasgupta, S.~Marzani, and L.~Tomlinson,
  \ifthenelse{\boolean{articletitles}}{\emph{{Predictions for Drell-Yan
  $\phi^*$ and $Q_T$ observables at the LHC}},
  }{}\href{https://doi.org/10.1016/j.physletb.2012.07.035}{Phys.\ Lett.\
  \textbf{B715} (2012) 152},
  \href{http://arxiv.org/abs/1205.4760}{{\normalfont\ttfamily
  arXiv:1205.4760}}\relax
\mciteBstWouldAddEndPuncttrue
\mciteSetBstMidEndSepPunct{\mcitedefaultmidpunct}
{\mcitedefaultendpunct}{\mcitedefaultseppunct}\relax
\EndOfBibitem
\bibitem{Alioli:2015toa}
S.~Alioli {\em et~al.}, \ifthenelse{\boolean{articletitles}}{\emph{{Drell-Yan
  production at $NNLL'+NNLO$ matched to parton showers}},
  }{}\href{https://doi.org/10.1103/PhysRevD.92.094020}{Phys.\ Rev.\
  \textbf{D92} (2015) 094020},
  \href{http://arxiv.org/abs/1508.01475}{{\normalfont\ttfamily
  arXiv:1508.01475}}\relax
\mciteBstWouldAddEndPuncttrue
\mciteSetBstMidEndSepPunct{\mcitedefaultmidpunct}
{\mcitedefaultendpunct}{\mcitedefaultseppunct}\relax
\EndOfBibitem
\bibitem{Coradeschi:2017zzw}
F.~Coradeschi and T.~Cridge,
  \ifthenelse{\boolean{articletitles}}{\emph{{reSolve -- A transverse momentum
  resummation tool}},
  }{}\href{https://doi.org/10.1016/j.cpc.2018.11.024}{Comput.\ Phys.\ Commun.\
  \textbf{238} (2019) 262},
  \href{http://arxiv.org/abs/1711.02083}{{\normalfont\ttfamily
  arXiv:1711.02083}}\relax
\mciteBstWouldAddEndPuncttrue
\mciteSetBstMidEndSepPunct{\mcitedefaultmidpunct}
{\mcitedefaultendpunct}{\mcitedefaultseppunct}\relax
\EndOfBibitem
\bibitem{Camarda:2019zyx}
S.~Camarda {\em et~al.}, \ifthenelse{\boolean{articletitles}}{\emph{{DYTurbo:
  Fast predictions for Drell-Yan processes}},
  }{}\href{http://arxiv.org/abs/1910.07049}{{\normalfont\ttfamily
  arXiv:1910.07049}}\relax
\mciteBstWouldAddEndPuncttrue
\mciteSetBstMidEndSepPunct{\mcitedefaultmidpunct}
{\mcitedefaultendpunct}{\mcitedefaultseppunct}\relax
\EndOfBibitem
\bibitem{Bizon:2018foh}
W.~Bizo\'{n} {\em et~al.}, \ifthenelse{\boolean{articletitles}}{\emph{{Fiducial
  distributions in Higgs and Drell-Yan production at N$^{\textit{3}}$\!LL +
  NNLO}}, }{}\href{https://doi.org/10.1007/JHEP12(2018)132}{JHEP \textbf{12}
  (2018) 132}, \href{http://arxiv.org/abs/1805.05916}{{\normalfont\ttfamily
  arXiv:1805.05916}}\relax
\mciteBstWouldAddEndPuncttrue
\mciteSetBstMidEndSepPunct{\mcitedefaultmidpunct}
{\mcitedefaultendpunct}{\mcitedefaultseppunct}\relax
\EndOfBibitem
\bibitem{Bellm:2015jjp}
J.~Bellm {\em et~al.}, \ifthenelse{\boolean{articletitles}}{\emph{{Herwig
  7.0/Herwig++ 3.0 release note}},
  }{}\href{https://doi.org/10.1140/epjc/s10052-016-4018-8}{Eur.\ Phys.\ J.\
  \textbf{C76} (2016) 196},
  \href{http://arxiv.org/abs/1512.01178}{{\normalfont\ttfamily
  arXiv:1512.01178}}\relax
\mciteBstWouldAddEndPuncttrue
\mciteSetBstMidEndSepPunct{\mcitedefaultmidpunct}
{\mcitedefaultendpunct}{\mcitedefaultseppunct}\relax
\EndOfBibitem
\bibitem{Sjostrand:2006za}
T.~Sj\"{o}strand, S.~Mrenna, and P.~Skands,
  \ifthenelse{\boolean{articletitles}}{\emph{{PYTHIA 6.4 physics and manual}},
  }{}\href{https://doi.org/10.1088/1126-6708/2006/05/026}{JHEP \textbf{05}
  (2006) 026}, \href{http://arxiv.org/abs/hep-ph/0603175}{{\normalfont\ttfamily
  arXiv:hep-ph/0603175}}\relax
\mciteBstWouldAddEndPuncttrue
\mciteSetBstMidEndSepPunct{\mcitedefaultmidpunct}
{\mcitedefaultendpunct}{\mcitedefaultseppunct}\relax
\EndOfBibitem
\bibitem{Sjostrand:2007gs}
T.~Sj\"{o}strand, S.~Mrenna, and P.~Skands,
  \ifthenelse{\boolean{articletitles}}{\emph{{A brief introduction to PYTHIA
  8.1}}, }{}\href{https://doi.org/10.1016/j.cpc.2008.01.036}{Comput.\ Phys.\
  Commun.\  \textbf{178} (2008) 852},
  \href{http://arxiv.org/abs/0710.3820}{{\normalfont\ttfamily
  arXiv:0710.3820}}\relax
\mciteBstWouldAddEndPuncttrue
\mciteSetBstMidEndSepPunct{\mcitedefaultmidpunct}
{\mcitedefaultendpunct}{\mcitedefaultseppunct}\relax
\EndOfBibitem
\bibitem{Gleisberg:2008ta}
T.~Gleisberg {\em et~al.}, \ifthenelse{\boolean{articletitles}}{\emph{{Event
  generation with SHERPA 1.1}},
  }{}\href{https://doi.org/10.1088/1126-6708/2009/02/007}{JHEP \textbf{02}
  (2009) 007}, \href{http://arxiv.org/abs/0811.4622}{{\normalfont\ttfamily
  arXiv:0811.4622}}\relax
\mciteBstWouldAddEndPuncttrue
\mciteSetBstMidEndSepPunct{\mcitedefaultmidpunct}
{\mcitedefaultendpunct}{\mcitedefaultseppunct}\relax
\EndOfBibitem
\bibitem{Skands:2010ak}
P.~Skands, \ifthenelse{\boolean{articletitles}}{\emph{{Tuning Monte Carlo
  generators: the Perugia tunes}},
  }{}\href{https://doi.org/10.1103/PhysRevD.82.074018}{Phys.\ Rev.\
  \textbf{D82} (2010) 074018},
  \href{http://arxiv.org/abs/1005.3457}{{\normalfont\ttfamily
  arXiv:1005.3457}}\relax
\mciteBstWouldAddEndPuncttrue
\mciteSetBstMidEndSepPunct{\mcitedefaultmidpunct}
{\mcitedefaultendpunct}{\mcitedefaultseppunct}\relax
\EndOfBibitem
\bibitem{Skands:2014pea}
P.~Skands, S.~Carrazza, and J.~Rojo,
  \ifthenelse{\boolean{articletitles}}{\emph{{Tuning PYTHIA 8.1: the Monash
  2013 tune}}, }{}\href{https://doi.org/10.1140/epjc/s10052-014-3024-y}{Eur.\
  Phys.\ J.\  \textbf{C74} (2014) 3024},
  \href{http://arxiv.org/abs/1404.5630}{{\normalfont\ttfamily
  arXiv:1404.5630}}\relax
\mciteBstWouldAddEndPuncttrue
\mciteSetBstMidEndSepPunct{\mcitedefaultmidpunct}
{\mcitedefaultendpunct}{\mcitedefaultseppunct}\relax
\EndOfBibitem
\bibitem{Aad:2014xaa}
ATLAS collaboration, G.~Aad {\em et~al.},
  \ifthenelse{\boolean{articletitles}}{\emph{{Measurement of the $Z/\gamma^*$
  boson transverse momentum distribution in $pp$ collisions at $\sqrt{s}$ = 7
  TeV with the ATLAS detector}},
  }{}\href{https://doi.org/10.1007/JHEP09(2014)145}{JHEP \textbf{09} (2014)
  145}, \href{http://arxiv.org/abs/1406.3660}{{\normalfont\ttfamily
  arXiv:1406.3660}}\relax
\mciteBstWouldAddEndPuncttrue
\mciteSetBstMidEndSepPunct{\mcitedefaultmidpunct}
{\mcitedefaultendpunct}{\mcitedefaultseppunct}\relax
\EndOfBibitem
\bibitem{Pietrulewicz:2017gxc}
P.~Pietrulewicz, D.~Samitz, A.~Spiering, and F.~J. Tackmann,
  \ifthenelse{\boolean{articletitles}}{\emph{{Factorization and resummation for
  massive quark effects in exclusive Drell-Yan}},
  }{}\href{https://doi.org/10.1007/JHEP08(2017)114}{JHEP \textbf{08} (2017)
  114}, \href{http://arxiv.org/abs/1703.09702}{{\normalfont\ttfamily
  arXiv:1703.09702}}\relax
\mciteBstWouldAddEndPuncttrue
\mciteSetBstMidEndSepPunct{\mcitedefaultmidpunct}
{\mcitedefaultendpunct}{\mcitedefaultseppunct}\relax
\EndOfBibitem
\bibitem{Bagnaschi:2018dnh}
E.~Bagnaschi, F.~Maltoni, A.~Vicini, and M.~Zaro,
  \ifthenelse{\boolean{articletitles}}{\emph{{Lepton-pair production in
  association with a $b\overline{b}$ pair and the determination of the $W$
  boson mass}}, }{}\href{https://doi.org/10.1007/JHEP07(2018)101}{JHEP
  \textbf{07} (2018) 101},
  \href{http://arxiv.org/abs/1803.04336}{{\normalfont\ttfamily
  arXiv:1803.04336}}\relax
\mciteBstWouldAddEndPuncttrue
\mciteSetBstMidEndSepPunct{\mcitedefaultmidpunct}
{\mcitedefaultendpunct}{\mcitedefaultseppunct}\relax
\EndOfBibitem
\bibitem{Bacchetta:2018lna}
A.~Bacchetta {\em et~al.}, \ifthenelse{\boolean{articletitles}}{\emph{{Effect
  of flavor-dependent partonic transverse momentum on the determination of the
  $W$ boson mass in hadronic collisions}},
  }{}\href{https://doi.org/10.1016/j.physletb.2018.11.002}{Phys.\ Lett.\
  \textbf{B788} (2019) 542},
  \href{http://arxiv.org/abs/1807.02101}{{\normalfont\ttfamily
  arXiv:1807.02101}}\relax
\mciteBstWouldAddEndPuncttrue
\mciteSetBstMidEndSepPunct{\mcitedefaultmidpunct}
{\mcitedefaultendpunct}{\mcitedefaultseppunct}\relax
\EndOfBibitem
\bibitem{ATL-PHYS-PUB-2014-015}
\ifthenelse{\boolean{articletitles}}{\emph{{Studies of theoretical
  uncertainties on the measurement of the mass of the $W$ boson at the LHC}},
  }{}  \href{http://cds.cern.ch/record/1956455}{ATL-PHYS-PUB-2014-015}, CERN,
  Geneva, 2014\relax
\mciteBstWouldAddEndPuncttrue
\mciteSetBstMidEndSepPunct{\mcitedefaultmidpunct}
{\mcitedefaultendpunct}{\mcitedefaultseppunct}\relax
\EndOfBibitem
\bibitem{Ball:2013hta}
NNPDF collaboration, R.~D. Ball {\em et~al.},
  \ifthenelse{\boolean{articletitles}}{\emph{{Parton distributions with QED
  corrections}},
  }{}\href{https://doi.org/10.1016/j.nuclphysb.2013.10.010}{Nucl.\ Phys.\
  \textbf{B877} (2013) 290},
  \href{http://arxiv.org/abs/1308.0598}{{\normalfont\ttfamily
  arXiv:1308.0598}}\relax
\mciteBstWouldAddEndPuncttrue
\mciteSetBstMidEndSepPunct{\mcitedefaultmidpunct}
{\mcitedefaultendpunct}{\mcitedefaultseppunct}\relax
\EndOfBibitem
\bibitem{Barlow:1993dm}
R.~J. Barlow and C.~Beeston,
  \ifthenelse{\boolean{articletitles}}{\emph{{Fitting using finite Monte Carlo
  samples}}, }{}\href{https://doi.org/10.1016/0010-4655(93)90005-W}{Comput.\
  Phys.\ Commun.\  \textbf{77} (1993) 219}\relax
\mciteBstWouldAddEndPuncttrue
\mciteSetBstMidEndSepPunct{\mcitedefaultmidpunct}
{\mcitedefaultendpunct}{\mcitedefaultseppunct}\relax
\EndOfBibitem
\bibitem{Conway:2011in}
J.~S. Conway, \ifthenelse{\boolean{articletitles}}{\emph{{Incorporating
  nuisance parameters in likelihoods for multisource spectra}},
  }{}\href{https://doi.org/10.5170/CERN-2011-006.115}{{PHYSTAT workshop on
  statistical issues related to discovery claims in search experiments and
  unfolding, CERN, Geneva, Switzerland} (2011) 115},
  \href{http://arxiv.org/abs/1103.0354}{{\normalfont\ttfamily
  arXiv:1103.0354}}\relax
\mciteBstWouldAddEndPuncttrue
\mciteSetBstMidEndSepPunct{\mcitedefaultmidpunct}
{\mcitedefaultendpunct}{\mcitedefaultseppunct}\relax
\EndOfBibitem
\bibitem{Eadie:100342}
W.~T. Eadie {\em et~al.}, {\em {Statistical methods in experimental physics}},
  North-Holland, Amsterdam, 1971\relax
\mciteBstWouldAddEndPuncttrue
\mciteSetBstMidEndSepPunct{\mcitedefaultmidpunct}
{\mcitedefaultendpunct}{\mcitedefaultseppunct}\relax
\EndOfBibitem
\end{mcitethebibliography}
\end{document}